\documentclass[twocolumn,showpacs,preprintnumbers,amsmath,amssymb]{revtex4}

\usepackage{graphicx}
\usepackage{dcolumn}
\usepackage{bm}

\begin{document}

\title{ Bogolyubov-Hartree-Fock approach
to studying the QCD ground state}

\author{S.V. Molodtsov}
 \altaffiliation[Also at ]{
Institute of Theoretical and Experimental Physics, Moscow, RUSSIA}
\affiliation{%
Joint Institute for Nuclear Research, Dubna,
Moscow region, RUSSIA
}%
\author{G.M. Zinovjev}
\affiliation{
Bogolyubov Institute for Theoretical Physics,
National Academy of Sciences of Ukraine, Kiev, UKRAINE
}%

\date{\today}

\begin{abstract}
 The quark's behaviour while influenced by a strong stochastic gluon field is
analyzed. An approximate procedure for calculating the effective Hamiltonian is
developed and the corresponding ground state within the Hartree-Fock-Bogolyubov approach
is found. The comparative analysis of various Hamiltonian models is given and
transition to the chiral limit in the Keldysh model is discussed in detail.
\end{abstract}

\maketitle

Nowadays we know that the  mixing of the zero modes is the microscopic mechanism
of the spontaneous breakdown of chiral symmetry in the instanton liquid model
\cite{ss}. In this approach the quarks are considered in a given gluon
background and the spectrum of the respective Dirac operator is calculated in order to be
accompanied then by averaging over the gluon ensemble. It is believed that at
low energy the zero modes are effectually overlapped and the eigenvalues of
Dirac operator spread over some range of virtualities.
In other words, studying the behaviour of a single quark in external
(stochastic) field one endeavours to guess the corresponding one-particle Green
function but for the quark ensemble even now. Being unable to carry out this transition in detail
from the first principles one is forced to devise the suitable approximations argued by
some general theoretical reasonings.
Also great care is needed in order to obtain proper
thermodynamical limit with nonzero chiral condensate. A lot of that happens to
be in striking contrast  to the Nambu-Jona-Lasinio (NJL) model \cite{4} which is
cognate to the instanton liquid model based actually on the similar multi-fermion
interaction. Superficially, the main distinction consists in the appearance of
some non-local formfactors instead of corresponding coupling constant.
As to the microscopic consideration, the generation of dynamical quark mass
in the NJL model is caused by the reconstruction of the Hamiltonian ground
state and the quarks manifest themselves already as the quasi-particles
\cite{NJL} although the multi-fermion attractive force should be strong enough, roughly
speaking. In this paper we emphasize that an instanton model and several other models which
are based on treating the stochastic ensemble of strong gluon field become practically
identical in many aspects to the NJL model.

Such an approach is motivated by the conceptual idea of an intricate nature of
the QCD vacuum \cite{CDG} having populated by intensive stochastic gluon fields
of nontrivial topological structure. Moreover, studying the corresponding cooled lattice
configurations gives evidence of this component presence \cite{cooled} and using the instantons
in the singular gauge to fit the data turns out to be very fruitful \cite{lattice} and allows one to
evaluate the ensemble density (around one topological charge per fm$^4$) and the
characteristic size of a saturating configuration (about $1$ GeV$^{-1}$).
Both estimates are in fairly good agreement with the
corresponding results of instanton liquid model \cite{DiakPetr}. Nevertheless,
the keen search of various confining configurations is still going on
\cite{Faddeev}, \cite{vanBaal}, \cite{Protog} in parallel with collecting the convincing
evidences that the construction of self-consistent ensemble of such configurations is a too
complicated problem (see, for example, the estimate for the (anti-)instanton ensemble done in Ref.
\cite{MolZin}).

Supposing the high-frequency component of stochastic ensemble of gluon fields as
the dominating contribution, we develop, in fact, an effective
theory\footnote{Which usually encodes the predictions of a quantum field theory
at low energies, and in which all assumptions done in the way to construct it
are not of special importance. What is entirely restrictive to fix the effective
action at really low energy (i.e. low cutoff) up to a few coupling
constants, to develop the approximate procedure to analyze the quark
interactions and to introduce the corresponding low energy effective variables
is an idea to neglect all the contributions coming from gluon fields generated
by the (anti-)quarks. Actually,it means the removal of corresponding cutoff(s)
from consideration but by the definition of an effective
theory this operation does not pose itself.}
with applying the procedure of simplified (averaged in time) system description
which is widely used at studying the dynamical systems.
Developing the effective theories which are discussed here has been launched to
a considerable extent by studying the behaviour of light quarks in the instanton
gas (liquid) \cite{GtH}. The zero mode approximation has provided for the quantitative
picture of spontaneous chiral symmetry breaking \cite{DPSb}. However, an effective
Lagrangian of the NJL type was soon received in Ref. \cite{Pobyl} by the direct
summation of certain leading diagrams and the obtained vertices of multi-quark
interactions occurred rather different from those calculated in the zero mode
approximation. Analysis of heavy quark systems behaviour affected by the stochastic
gluon fields \cite{DS} has demonstrated that at constructing the respective effective
theory the cluster decomposition of generating functional \cite{2} can be very
efficient tool and the specific role of various characteristic correlation times
has been clarified to classify the descriptions. These results together with the
criticism of zero mode approximation \cite{Ker} have contributed to widening the
cluster decomposition applications. This approach has been used to analyse the
light quark behaviour \cite{Simon} and it is interesting to note
the effective Lagrangian has agreed with that obtained in \cite{Pobyl}.
In the context of our interest here the cluster decomposition is called upon to
describe the correlations in quantum system inspired by an external process. In
this situation, as a matter of fact, the description of system behaviour is executed
by averaging the generating functional. However, as we show in this paper such a
procedure applied to the quantum system could be incomplete and it is more
appropriate to base an analysis on the corresponding density matrix. Nonetheless,
we argue here that in the 'white noise' limit (when the time intervals of
stochastic impulses are very short) the procedure of averaging the generating
functional occurs quite adequate.

The form of the effective Hamiltonian obtained urges us to search the system
ground state as the Bogolyubov trail function. The corresponding dressing
transformation will be analysed for various formfactors of effective Hamiltonian.
In such an approach the quarks are already treated as the quasi-particles and rather
practical way to get beyond the zero mode approximation appears. It grounds on the
method of simple iterations of corresponding integral equation solutions for the
dressing transformation which quite stable unlike the results of mean field
approximation \cite{Pobyl}, \cite{Simon}. The different ensembles are examined and
their selection is stipulated by the requirement that one of their asymptotic forms
would be the NJL model which plays a calibrating role in our calculations. The chiral
limit of the Keldysh model with the correlator behaving as a $\delta$-function in the
momentum space is studied in detail and the singular behaviour of
the corresponding mean energy functional is demonstrated.

\section{The Hartree--Fock--Bogolyubov approximation}
We consider the quark (anti-quark) ensemble in the background of strong
stochastic gluon field and suppose this field is so strong that we could neglect
the gluon interchanging processes (quenched approximation). The stochastic gluon
field is characterized by a correlation function and its particular form will be
discussed and fixed below. The Lagrangian density is the following
\begin{equation}
\label{1}
{\cal L}_E=\bar q~(i\gamma_\mu D_\mu+im)~q~,
\end{equation}
here  $q$, $\bar q$ are the quark and anti-quark fields with covariant
derivative $D_\mu=\partial_\mu -i g A^a_\mu t^a$ where $A^a_\mu$ is the  gluon
field, $t^a=\lambda^a/2$ are the generators of colour gauge group $SU(N_c)$ and
$m$ is the current quark mass, $\mu=1,2,3,4$. We work in the context of the Euclidean
field theory and $\gamma_\mu$ mean the Hermitian Dirac matrices
($\gamma^{+}_\mu=\gamma_\mu$, $\{\gamma_\mu,\gamma_\nu\}=2~\delta_{\mu\nu}$)
in the chiral representation. Then the corresponding Hamiltonian description results from
\begin{equation}
\label{2}
{\cal H}=\pi \dot q-{\cal L}_E~,~~\pi=\frac{\partial{\cal L}_E}{\partial \dot
q}=i q^{+}~,
\end{equation}
and, in particular, for the noninteracting fields we have
\begin{equation}
\label{3}
{\cal H}_0=-\bar q~(i{\bf \gamma}{\bf \nabla}+im)~q~.
\end{equation}
In the Schr\"odinger representation the quark field evolution is determined by
the equation for the quark probability amplitude $\Psi$ as
\begin{equation}
\label{4}
\dot \Psi=- H \Psi~,
\end{equation}
and the creation and annihilation operators of quarks and anti-quarks $a^+, a$,
$b^+, b$ have no 'time' dependence and consequently look like
\begin{widetext}
\begin{equation}
\label{5}
q_{\alpha i}({\bf x})=\int\frac{d {\bf p}}{(2\pi)^3} \frac{1}{(2|p_4|)^{1/2}}~
\left[~a({\bf p},s,c)~u_{\alpha i}({\bf p},s,c)~ e^{i{\bf p}{\bf x}}+
b^+({\bf p},s,c)~v_{\alpha i}({\bf p},s,c)~ e^{-i{\bf p}{\bf x}}\right]~,
\end{equation}
\end{widetext}
here the summation over index $s$ which stands to describe two quark spin
polarizations and index $c$ which should play the similar role for a colour is
implied. Further we make concrete the form of the Dirac conjugated spinor.
Fixing a spin polarization as it is known can be done by imposing an additional
constraint on spinor (see, below). However, there is no direct analogy with
the colour polarization and the particular state should be fixed by the
corresponding complete set of diagonal operators which includes the Casimir
operators as well. In fact, this complete definition of the spinor colour state
is unnecessary for us here. All observables are usually expressed by summing up
the polarization states of some bilinear spinor combinations as the singlet and
octet states and the singlet component is obviously playing the specific role.

The density of interaction Hamiltonian can be presented as
\begin{equation}
\label{6}
{\cal V}_S=\bar q({\bf x})~t^a\gamma_\mu A^{a}_\mu(t,{\bf x})~q({\bf x})~.
\end{equation}
The obvious dependence on 'time' in this Hamiltonian is present in the gluon
field only. As it is mentioned above we are planning to work with the stochastic
gluon field implying the random process for which one may define only a
probability of realizing some gluon configuration. Such a nature of gluon field
urges (and allows) us to develop the approximate procedure for describing the
quark field treating (\ref{4}) as a probabilistic process. Then the system states
are described by the corresponding averages (over a 'time' or an ensemble according
to the ergodic hypothesis). However, in the quantum theory we face one difficulty
in this way because $\Psi$ is a probability amplitude and an immediate averaging
of $\langle\Psi \rangle$ can be insignificant. Studying a mean probability density
$\langle\stackrel{*}{\Psi}\stackrel{}{\Psi} \rangle$ looks more promising and can
be realized by complicating the procedure of continual integration \cite{1}.
In Appendix I we analyse convincing quantum mechanical example to illustrate the
difference between two approaches. One of those is based on constructing the
corresponding density matrix $\langle \stackrel{*}{\Psi}\stackrel{}{\Psi}\rangle$,
and the second approach does use the relevant averaging of the functional
$\langle \Psi \rangle$. We argue the latter could be practical for application in
the 'white noise' limit with the $\delta$-like time correlation function.
Adapting these ideas to the gauge theories we should obviously strive to operate
with the gauge invariant quantities which include an ordered exponential, at
least. Unfortunately, such a program in what concerns the ensemble consideration
is still very far to be realized. However, it is clear that applying the averaging
procedure would result in putting in an appearance of a set of
corresponding correlation functions $\langle A^2\rangle$, $\langle A^{4}\rangle$
etc.\footnote{For example, the spontaneous breaking of chiral symmetry is well
understood in the instanton liquid model just due to such a trick \cite{Pobyl},
\cite{Simon}. It is interesting to notice here then correlation function series
summed up is expressed in the highest order of the packing fraction
parameter $n\bar \rho^{4}$, where $n$ is the instanton liquid density and
$\bar\rho$ is the mean size of (anti-)instanton, with the covariant derivative in the
field of each separate (anti-)instanton and includes also the free Green function.}

In the interaction representation, where $\Psi=e^{H_0 t}\Phi$, Eq.(\ref{4}) can
be rewritten as
\begin{equation}
\label{7}
\dot \Phi=- V \Phi~,~~V=e^{H_0 t} V_S e^{-H_0 t}~.
\end{equation}
Now the 'time' dependence appears in quark operators as well.
Now we remind some features of the averaging description as formulated in
Ref.\cite{2}. Presenting Eq.(\ref{7}) in the integral form as
\begin{equation}
\label{7n}
\Phi(t)= \Phi(0)-\int^t_0~d\tau~ V(\tau)~\Phi(\tau)~,
\end{equation}
where $\Phi(0)$ is an arbitrary initial state of ensemble and performing another
iteration one receives
$$
\Phi(t)= \Phi(0)-\int^t_0\!\! d\tau V(\tau)\Phi(0)+\int^t_0\!\! d\tau V(\tau)
\!\!\int^\tau_0 \!\!d\tau' V(\tau') \Phi(\tau').
$$
By averaging the fast-changing component and uncoupling the correlators one
approximately approaches the long wavelength component $\langle \Phi \rangle$ in
the highest order (also taking into account that $\langle V\rangle=0$) as
follows
\begin{equation}
\label{7n2}
\langle \Phi(t)\rangle\approx \Phi(0)+\int^t_0 \!\!d\tau \int^\tau_0\!\! d\tau'
\langle
V(\tau)V(\tau')\rangle
\langle\Phi(\tau')\rangle.
\end{equation}
Certainly, it is assumed the characteristic correlation time of stochastic
process is smaller than the time characteristic for the process
$\langle \Phi\rangle$. By differentiating Eq.(\ref{7n2}) it is easy to get rid of the initial
condition $\Phi(0)$ and to have
$$
\langle \dot\Phi(t)\rangle=+\int^t_0~d\tau'~\langle V(t)V(\tau')\rangle~
\langle\Phi(\tau')\rangle~.
$$
Actually this equation should describe a steady-state process
and at reversing a time the solution, in general case, will not return to the
initial magnitude  $\Phi(0)$. Changing the integration variable as
$\tau'=t-\tau$ one comes to
\begin{equation}
\label{7n4}
\langle \dot\Phi\rangle=+\int^t_0~d\tau~\langle V(t)V(t-\tau)\rangle~
\langle\Phi(t-\tau)\rangle~.
\end{equation}
It is usually supposed the correlations are quickly decaying then the upper
limit of integration might be changed for $\infty$ and in order to deal with the
local process it is well justified (without a precision loss) to change the argument
of function $\langle \Phi \rangle$ for $t$. Eventually, as a result we have
\begin{equation}
\label{8}
\langle\dot\Phi(t)\rangle=+\int_0^{\infty} d\tau ~\langle V(t) V(t-\tau)
\rangle~\langle\Phi(t)\rangle~.
\end{equation}
(The requirements to validate the factorization of the long-wavelength component
are discussed, for example, in \cite{2}.) Implementing the approximation (\ref{8})
in the quantum field theory models, we run into the trouble at trying to get the
most general form of correlation function if the characteristic quark and gluon
correlation times are comparable. Fortunately, if the quark fields are
considered to be practically constant on the gluon background the problem
receives essential simplification. The gluon field contribution may be factorized as
a corresponding correlation function
$\langle A^a_\mu(x)A^b_\nu(y)\rangle$ \cite{7}.
Recent lattice measurements of this correlation function provide us with a
reasonable arguments to interpret the result as gluon 'mass' generation
($\sim 300$ -- $400$ MeV) in the momentum region of order $200$ MeV
\cite{Bornyak}.

It is curious to notice that the averaging over ensemble ('time') in the
right hand side of Eq.(\ref{8}) is performed in both the correlator and
$\langle\Phi(t)\rangle$. It means that by resumming and averaging a certain
class of diagrams in the quantum field theory models, one may take into account
high order correlator contributions in different ways if the form of function
$\langle\Phi(t)\rangle$ is specified. Besides, the correlation functions in models
interesting to us should be translation invariant and it implies that the
correlator in Eq.(\ref{8}) has the following form
$$\langle V(t) V(t-\tau)\rangle=F(\tau)~,$$
i.e., for example, an one-dimensional process after having done the integration
in Eq.(\ref{8}) will be described by a constant which characterizes the slow
process. In quantum field theory for the problem we are interested in,
the correlator connecting two space  points
\begin{widetext}
$$
\langle\dot\Phi(t)\rangle=\int d{\bf x}~ \bar q({\bf x},t)~
t^a\gamma_\mu~q({\bf x},t)~
\int_0^{\infty} d\tau \int d{\bf y} ~\bar q({\bf y},t-\tau)~
t^b\gamma_\nu~q({\bf y},t-\tau)~
g^2\langle A^{a}_\mu(t,{\bf x}) A^{b}_\nu(t-\tau,{\bf y})\rangle~
\langle\Phi(t)\rangle~$$
\end{widetext}
appears instead of a constant. Assuming the correlation function is rapidly
decreasing with time we change the 'time' $t-\tau$ dependence in the quark
fields for $t$ and perform the inverse transformation to the Schr\"odinger
representation. Then introducing the function $\chi=e^{-H_0 t}\langle\Phi\rangle$
we have{\footnote{Let us notice that in general this form does not coincide with
$\langle e^{-H_0 t}\Phi\rangle$.}} the following equation
\begin{widetext}
\begin{eqnarray}
\label{9}
&&\dot \chi=- H_{ind}~ \chi~,\nonumber\\ [-.2cm]
\\ [-.25cm]
&&{\cal H}_{ind}=-\bar q~(i{\bf \gamma}{\bf \nabla}+im)~q-\bar
q~t^a\gamma_\mu~q~
  \int d{\bf y} ~\bar q'~t^b\gamma_\nu~q'~
\int_0^{\infty} d\tau~ g^2\langle A^{a}_\mu A^{'b}_\nu\rangle~,\nonumber
\end{eqnarray}
\end{widetext}
where $q=q({\bf x})$, $\bar q=\bar q({\bf x})$,
$q'=q({\bf y})$, $\bar q'=\bar q({\bf y})$,
$A^{a}_\mu =A^{a}_\mu(t,{\bf x})$ and $A^{'b}_\nu=A^{b}_\nu(t-\tau,{\bf y})$.

In order to receive the final result we should fix the form of correlation
function. In this paper we rely on the stochastic ensemble of (anti-)instantons
in the singular gauge
\begin{equation}
\label{anz}
A^{a}_\mu(x)=\sum_{i=1}^N~A^{a}_\mu(x;\gamma_i)~,
\end{equation}
and instanton solution reads as
\begin{eqnarray}
\label{23}
&&A^{a}_\mu(x)=\frac{2}{g} 4\pi^2 i\rho^2 \omega^{ab} \bar\eta_{\mu b\nu}\int
\frac{dq}{(2\pi)^4}~q_\nu~\phi(q)~e^{iq(x-z)}~,\nonumber\\[-.2cm]
\\ [-.25cm]
&&\phi(q)=\frac{1}{q^2}~\left(K_2(q\rho)-
\frac{2}{q^2\rho^2}\right)~,\nonumber
\end{eqnarray}
where $K_2$ is the modified Bessel function of imaginary argument, $\rho$ is the
instanton size, the matrix $\omega$ appoints the pseudo-particle orientation in
colour space, $z$ is the coordinate of instanton center and $\bar\eta$ stands
for the 't Hooft symbol. The distribution of the pseudo-particle orientation in
colour space is supposed to be homogeneous $\sim d\omega$ as well as the probability
to find a pseudo-particle in the volume element is proportional $\sim dz/V$ where
$V$ is the volume of the system under consideration. Apparently, specifying the
saturating configuration in the form of Eq.(\ref{anz}) is, in a direct way, the
gauge fixing procedure. Calculating the quantum corrections for every single
pseudo-particle in one-loop approximation (what corresponds to the zeroth order of
the $N/V$-expansion), and exploiting the variation principle \cite{DiakPetr},
\cite{MolZin} allows one to ascertain the size distribution of pseudo-particles.
In this way it is possible to attach clear meaning to the functional and to construct
in the thermodynamical limit $\lim_{V\to \infty} N/V\to n$ the state possessing a
negative energy density and developing a non-zero gluon condensate.
(Uncertain interrelation of perturbative and non-perturbative contributions into
the path integral \cite{shafer} makes the computability of generating functional
highly nontrivial as for now.) In Eq.(\ref{9}) we imply the correlation function
integrated over the 'time' for which we receive in the highest order in the density
$n$ of (anti-)instanton ensemble
\begin{eqnarray}
\label{24}
&&\int_0^{\infty} d x_4 ~\langle A^{a}_\mu(x) A^{b}_\nu(y)\rangle=\frac{1}{2}
\int_{-\infty}^{\infty} d x_4~ \langle A^{a}_\mu(x) A^{b}_\nu(y)\rangle=
\nonumber\\
&&=\frac{4(4\pi^2)^2}{g^2}~\frac{\delta_{ab}~n\rho^4}{N_c^2-
1}~(\delta_{\mu\nu}\delta_{\alpha\beta}-
\delta_{\mu\alpha}\delta_{\nu\beta})\times\nonumber\\
&&\times\int \frac{dp}{(2\pi)^4}~p_\alpha p_\beta~e^{ip(x-y)}~
\phi(-p)\phi(p)~\frac{1}{2}~2\pi~\delta(p_4)~.
\nonumber
\end{eqnarray}
The first equality is valid due to the symmetry properties of instanton solution.
Then the correlation function can be presented as
\begin{eqnarray}
\label{25}
&&\langle \widetilde{A^{a}_\mu A^{b}_\nu}({\bf p})\rangle=
\frac{(4\pi^2)^2~n\rho^4}{g^2}~\frac{2~\delta_{ab}}{N_c^2-
1}~\left[I(p)~\delta_{\mu\nu}-J_{\mu\nu}(p)\right]\nonumber\\ [-.2cm]
\\ [-.25cm]
&&I(p)={\bf p}^2~\phi(-p)\phi(p)~,~~
J_{ij}(p)= p_i p_j~\phi(-p)\phi(p)~,\nonumber\\
&&J_{4i}=J_{i4}=J_{44}=0~.\nonumber
\nonumber
\end{eqnarray}
We suppose in what follows the various stochastic ensembles of gluon fields are
characterized by their profile functions $I(p)$, $J_{\mu\nu}(p)$ and analyze the
contribution of quadratic correlator only. However, this deficiency of fixing the gauge
implicitly for the truncated system is compensated, in a sense, by our investigation of
full spectrum of reasonable correlation functions (including an opposite limiting
correlators when they are extrapolated even into the perturbative region).
Recent considerable progress in studying the confining configurations of lattice
gauge theories, in particular, revealing the monopole clusters and their role in
confinement (see, for review \cite{Zah}) as well as detecting the specific features of
quark behaviour in the uncooled configurations and the indications that low-lying Dirac
eigenmodes are localized on the objects of dimension inherent in monopoles and vortices
\cite{shrink} looks entirely urging. But these results are also bringing the perilous
tendencies because reveal the some features of lattice gauge theories common with the
compact electrodynamics. This fact makes almost inevitable to draw in the singular
(in the continual limit) objects and to give them an underlying physical meaning.
Searching the formfactors (the corresponding ensembles of saturating configurations)
interesting for applications one should compare to the reasonable results for the
four nonets of light mesons obtained in the NJL model. Apparently, the constants
of effective four-quark Hamiltonian (scalar, pseudo-scalar, vector and axial-vector
channels) and the parameters of integral saturation (cut-off) should be comparable.
Seems, it might be carefully supposed that those singular objects (at still an
unknown scale) have to reproduce the major features of successful effective NJL
Hamiltonian after the corresponding averaging. In any case, the problems to find the
specific features of such singular objects which admit their experimental
identification and to analyse the quark behaviour in the ensembles of monopoles or
vortices are of really great interest \cite{mm}.

With such a form of induced four-fermion interaction we are going to search the
ground state as the Bogolyubov probe function with vacuum quantum
numbers{\footnote{In order to avoid any misunderstanding we remind here that
fixing a form of ground state introduces a primary frame.}}  \cite{3}, \cite{Y}
\begin{widetext}
\begin{eqnarray}
\label{10}
&&|\sigma\rangle=T~|0\rangle~,\nonumber\\ [-.2cm]
\\ [-.25cm]
&&
T=\Pi_{ p,s,c}~\exp\{~\varphi~[~a^+({\bf p},s,c)~b^+(-{\bf p},s,c)+
a({\bf p},s,c)~b(-{\bf p},s,c)~]~\}~,\nonumber
\end{eqnarray}
\end{widetext}
which is defined by minimizing mean energy
\begin{equation}
\label{11}
E=\langle\sigma|H|\sigma\rangle~,
\end{equation}
here $\varphi=\varphi({\bf p})$ and $|0\rangle$ is the vacuum of free
Hamiltonian, i.e. $a({\bf p},s,c)~|0\rangle=0$, $b({\bf p},s,c)~|0\rangle=0$.
Introducing with the dressing $T$ transformation the creation and
annihilation operators of quasi-particles ($T^{-1}=T^{\dagger}$ for fermions)
$$A=T~a~T^{-1}~,~~~B^+=T~b^+T^{-1}~,$$
we present the operator Eq.(\ref{5}) as, with the Dirac conjugate spinor
\begin{widetext}
\begin{eqnarray}
\label{12}
&&q({\bf x})=\int\frac{d {\bf p}}{(2\pi)^3} \frac{1}{(2|p_4|)^{1/2}}~
\left[~A({\bf p},s,c)~U({\bf p},s,c)~e^{i{\bf p}{\bf x}}+
B^+({\bf p},s,c)~V({\bf p},s,c)~ e^{-i{\bf p}{\bf x}}\right]~,\nonumber\\[-.2cm]
\\ [-.25cm]
&&\bar q({\bf x})=\int\frac{d {\bf p}}{(2\pi)^3} \frac{1}{(2|p_4|)^{1/2}}~
\left[~A^+({\bf p},s,c)~\overline{U}({\bf p},s,c)~e^{-i{\bf p}{\bf x}}+
B({\bf p},s,c)~\overline{V}({\bf p},s,c)~ e^{i{\bf p}{\bf x}}\right]~,\nonumber
\end{eqnarray}
\end{widetext}
where the spinors $U$ and $V$ are defined as
\begin{eqnarray}
\label{13}
\!\!\!\!\!\!\!U({\bf p},s,c)&=&\cos(\varphi)~u({\bf p},s,c)-\sin(\varphi)~v(-
{\bf p},s,c)~,\nonumber\\ [-.2cm]
\\ [-.25cm]
\!\!\!\!\!\!\!V({\bf p},s,c)&=&\sin(\varphi)~u(-{\bf
p},s,c)+\cos(\varphi)~v({\bf p},s,c)~.\nonumber
\end{eqnarray}
with $\overline{U}({\bf p},s,c)=U^+({\bf p},s,c)~\gamma_4$ and
$\overline{V}({\bf p},s,c)=V^+({\bf p},s,c)~\gamma_4$.
Now we have to specify the choice of spinors in the Euclidean variables. They
obey the Dirac equations
\begin{equation}
\label{15}
(\hat p-im)~ u(p,s)=0~,~~~(\hat p+im)~ v(p,s)=0~,
\end{equation}
(with $\hat p=p_4\gamma_4 +{\bf p}{\bf \gamma}$) and additional constraint which
fixes the spinor
polarization
\begin{equation}
\label{16}
i\gamma_5~\hat s~ u(p,s)=u(p,s)~,~~~i\gamma_5~\hat s~ v(p,s)=v(p,s)~,
\end{equation}
where $\gamma_5 =-\gamma_1\gamma_2 \gamma_3 \gamma_4$, and the four-vector
$s$ is normalized to unit and orthogonal to the four-vector $p$, i.e. $s^2=1$,
$(ps)=0$.
It could be, for example,
$$s_4=\frac{({\bf p}{\bf n})}{im}~,~~~{\bf s}={\bf n}+\frac{({\bf p}{\bf n})
~{\bf p}}{im~(p_4-im)}~,$$
where ${\bf n}$ is an arbitrary unit vector. If the covariant normalization
conditions are satisfied
\begin{equation}
\label{17}
\bar u u=2im~,~~~\bar v v=-2im~,
\end{equation}
the spinors are defined with the precision up to phase factor. All these
conditions allow us to formulate the following matrix representation
\begin{eqnarray}
\label{18}
&&u(p,s)~\bar u(p,s)= \frac{\hat p+im}{2}~(1+i\gamma_5~\hat s),~~\nonumber\\[-
.2cm]
\\ [-.25cm]
&&v(p,s)~\bar v(p,s)= \frac{\hat p-im}{2}~(1+i\gamma_5~\hat s)~.\nonumber
\end{eqnarray}
Calculating the mean energy Eq.(\ref{11}) we meet spinors with opposite moments.
We introduce the four-vector $q=(p_4,-{\bf p})$ in order to simplify notations.
Using the projection operator we can express the spinor $v(q,s)$ through the
spinor $u(p,s)$ (see \cite{kriv})
\begin{equation}
\label{19}
v(q,s) =\alpha~\frac{\hat q-im}{-2im}~\frac{1+i\gamma_5~\hat s}{2}~u(p,s)~.
\end{equation}
The coefficient $\alpha$ is fixed by the covariant normalization Eq.(\ref{17})
up to the phase factor as
$$
\stackrel{*}{\alpha}\stackrel{}{\alpha}=-\frac{2~m^2}{(pq)+m^2}=\frac{m^2}{{\bf
p}^2}~,~~~
|\alpha|=\frac{m}{|{\bf p}|}~.$$
Then summing up over the spinor states results in
\begin{widetext}
\begin{eqnarray}
\label{20}
&&\sum_s u(q,s)~ \bar v(p,s)= ~\alpha ~\frac{\hat q+im}{2im}~(\hat p-im)~,~~
\sum_s v(p,s) ~\bar u(q,s)= ~\stackrel{*}{\alpha} ~(\hat p-im)~\frac{\hat
q+im}{2im}~,\nonumber\\ [-.2cm]
\\ [-.25cm]
&&
\sum_s u(p,s)~ \bar v(q,s)= ~\stackrel{*}{\alpha} ~(\hat p+im) ~\frac{\hat q-
im}{2im}~,~~
\sum_s v(q,s)~ \bar u(p,s)= ~\alpha~ \frac{\hat q-im}{2im}~(\hat
p+im)~.\nonumber
\end{eqnarray}
\end{widetext}
The polarization in which the momentum $\bf p$ and unit polarization vector $\bf
n$ are orthogonal $({\bf p}{\bf n})=0$ turns out to be the most convenient for
handling. In such a situation both operators $\hat p$ and $\hat q$ commute with
$\gamma_5 \hat s$ and the polarization directions of quark and anti-quark could
be taken identical (although in general case they should be two different
directions). Then the summation over polarization of quarks and anti-quarks is
performed separately in the final equations. It allows us not to control the
obligatory constraint to have the vacuum quantum numbers of the pairs present in
the intermediate calculations.

When calculating the mean energy Eq.(\ref{11}) nontrivial contribution bilinear
in quark operators comes from the terms of type
$\langle\sigma| B~B^+|\sigma\rangle$ (remember $B |\sigma\rangle=0$,
$A|\sigma\rangle=0$). The contribution from the terms like
$\langle\sigma| A~A^+|\sigma\rangle$ is absent because of the particular
representation of bilocal operator we are using as $\bar q q$ (then quadratic terms
are expressed by the spinors $V,\bar V$. Due to the similar reasons the four-quark
operators develop only two nonzero contributions
$\langle\sigma|B~B^+~B'~B^{'+}|\sigma\rangle$ and $\langle\sigma|
B~A~A^{'+}~B^{'+}|\sigma\rangle$. The first combination corresponds to the contribution
of so-called tadpole diagrams  and the latter is related to the
asterisk{\footnote{If one is interested in stochastic gluon field
contribution in the highest one-particle order approximation only all gluon
lines in the respective Feynman diagrams are depicted as coming from the single
point corresponding to the center coordinate of the gluon configuration. However,
working within the Green function method to derive the Schwinger-Dyson
equations suppose the resummation of special classes of diagrams. Then taking
into account the diagrams with a large number of asterisks (the centers of gluon
configurations) looks like an overrange of one-particle approximation. Thus the second
star coming to play, for example, signals the contribution of second order in density
available.}} diagrams. As a result the four-fermion interaction contribution can be
presented in the following form
\begin{widetext}
\begin{eqnarray}
\label{21}
&&\langle \widetilde{A^{a}_\mu A^{b}_\nu}(0)\rangle~\int\frac{d {\bf p}~
d {\bf p}'}{(2\pi)^6}\frac{1}{4~|p_4||p'_4|}~
\overline{V}_{\alpha i}({\bf p},s,c)~t^a_{ij}\gamma^\mu_{\alpha \beta}
V_{\beta j}({\bf p},s,c)~
\overline{V}_{\gamma k}({\bf p}',s',c')~t^b_{kl}\gamma^\mu_{\gamma
\delta}V_{\delta l}({\bf p}',s',c')+
\nonumber\\
&&+\int\frac{d {\bf p}~d {\bf p}'}{(2\pi)^6}
\frac{1}{4~|p_4||p'_4|}~
\overline{V}_{\alpha i}({\bf p},s,c)~t^a_{ij}\gamma^\mu_{\alpha \beta}
V_{\delta l}({\bf p},s,c)~
\overline{U}_{\gamma k}({\bf p}',s',c')~t^b_{kl}\gamma^\nu_{\gamma \delta}
U_{\beta j}({\bf p}',s',c')\langle \widetilde{A^{a}_\mu A^{b}_\nu}
({\bf p}+{\bf p}')\rangle.\nonumber
\end{eqnarray}
\end{widetext}
Here $\langle\widetilde{A^{a}_\mu A^{'b}}_\nu\rangle$ is the Fourier transform
of the gluon correlator and the summation over spinor and colour ~indices is
implied. The contribution of the first tadpole diagram is an identical zero due
to completeness of the spinor basis in color space, giving a unit color matrix
(in particular it is valid for colour singlet quark configuration). In
electrodynamics the considered  term provides a dominant contribution.
But it is interesting to remark that the singular character of
photon propagator in the infrared region makes this abelian theory even more
complicated to research than in the nonabelian one. In the compact $U(1)$
electrodynamics (on a lattice) the infrared behaviour of correlation function is
formed by the monopole contributions but nowadays it is still difficult to
define a scale where these effects show up themselves. In the octet channel of
nonabelian theory we obtain the quark repulsion $\sim -1/(4N_c)$ and therefore
this regime might be omitted when searching the minimum of mean energy Eq.(\ref{11}).
Then for the spinors with polarizations summed up we have
\begin{widetext}
\begin{eqnarray}
\label{22}
&&
V\overline{V}=p_4\gamma_4+\cos(\theta)~({\bf p}{\bf \gamma}-im)-
\frac{\stackrel{*}{\alpha}+\stackrel{}{\alpha}}{2im}~\sin(\theta)~
({\bf p}^2-im~{\bf p}{\bf \gamma})~,
\nonumber\\
&&U\overline{U}=p_4\gamma_4+\cos(\theta)~({\bf p}{\bf \gamma}+im)+
\frac{\stackrel{*}{\alpha}+\stackrel{}{\alpha}}{2im}~\sin(\theta)~
({\bf p}^2+im~{\bf p}{\bf \gamma})~,\nonumber
\end{eqnarray}
\end{widetext}
where angle $\theta=2\varphi$. In the formulae above the phase inherent in the
sum ${\stackrel{*}{\alpha}+\stackrel{}{\alpha}}$ (a spinor is defined up to such
a phase) is still indefinite. The direct analysis of the mean energy functional
demonstrates that the most preferable value of the phase factor (responsible for
the colour interaction of quarks) is the value when the coefficient $\alpha$ appears to be
a real number. For definiteness we put $\alpha=+|m|/p$. The curious fact is that the
results of summation are not equal ($V\overline{V}(m)=U\overline{U}(-m)$)
and they coincide in the chiral limit $m=0$ only, i.e. particles and
antiparticles formally generate the different contributions.

The direct calculations lead to the following result for the mean energy (\ref{11})
\begin{widetext}
\begin{eqnarray}
\label{26}
&&\langle\sigma|H_{ind}|\sigma\rangle=-\int \frac{d {\bf p}}{(2\pi)^3}~
\frac{2N_c~p_4^{2}}{|p_4|}\left(1-\cos\theta\right)-
\nonumber\\
&&-\widetilde G\int \frac{d {\bf p}d {\bf q}}{(2\pi)^6}\left\{-(3\widetilde I-
\widetilde J)
\frac{p_4~ q_4}{|p_4||q_4|}+(4\widetilde I-\widetilde J)\frac{p~ q}{|p_4||q_4|}
\left(\sin\theta-\frac{m}{p}\cos\theta\right)
\left(\sin\theta'-\frac{m}{q}\cos\theta'\right)-\right.\nonumber\\ [-.2cm]
\\ [-.25cm]
&&-\left.(2\widetilde I\delta_{ij}+2\widetilde J_{ij}-\widetilde
J\delta_{ij})~\frac{p_i~q_j}{|p_4||q_4|}~
\left(\cos\theta+\frac{m}{p}\sin\theta\right)
\left(\cos\theta'+\frac{m}{ q}\sin\theta'\right)~\right\}~,\nonumber
\end{eqnarray}
\end{widetext}
here we designated $p=|{\bf p}|$, $q=|{\bf q}|$,
$\widetilde I=\widetilde I({\bf p}+{\bf q})$,
$\widetilde J_{ij}=\widetilde J_{ij}({\bf p}+{\bf q})$,
$\widetilde J=\sum_{i=1}^3\widetilde J_{ii}$,
$p_4^2+p^2=q_4^2+q^2=-m^2$, $\theta'=\theta(q)$,
$\widetilde G=(4\pi^2)^2~n\rho^4$ and as a matter of convenience we singled out
the colour factor $G'=\frac{2}{N_c^{2}-1}\widetilde G$. To obtain this result we
performed the regularization (subtracting the free Hamiltonian $H_0$). It results
in the presence of a unit (together with $-\cos\theta$) in the
parentheses of the first integral. Let us  also remind that in the Euclidean
space $p_4^2$ is a negative magnitude. Then with Eq.(\ref{26}) available one can find
the most advantageous value of the angle $\theta$ from the condition
\begin{equation}
\label{27}
\frac{d\langle\sigma|H_{ind}|\sigma\rangle}{d\theta}=0~.
\end{equation}
Henceforth we characterize the different stochastic ensembles of the gluon
fields by their profile functions $I(p)$, $J_{\mu\nu}(p)$.

\section{Nambu-Jona-Lasinio model}
Now let us consider the example in which the correlation function behaves in the
coordinate space as the $\delta$-function (simply we assume $J_{\mu\nu}(p)=0$).
Actually, it corresponds to the Nambu-Jona-Lasinio (NJL) model \cite{4}. As well
known the regularization is required to obtain an intelligent result in this model.
We adjust the NJL model with the parameter set given by Ref. \cite{5}, and limit the
integration interval over momentum in Eq.(\ref{26}) with the quantity
$|{\bf p}|<\Lambda$ ($\Lambda=631$ MeV). Then the functional (\ref{26})
is written in the following form (unessential terms contributing the constant
values are omitted)
\begin{widetext}
\begin{equation}
\label{28}
W=\int^\Lambda \frac{d{\bf p}}{(2\pi)^3}~\left[|p_4|\left(1-\cos\theta\right)-
G\frac{p}{|p_4|}\left(\sin\theta-\frac{m}{p}\cos\theta\right)\int^\Lambda
\frac{d{\bf q}}{(2\pi)^3}
\frac{q}{|q_4|}\left(\sin\theta'-\frac{m}{q}\cos\theta'\right)\right]~,
\end{equation}
\end{widetext}
here $m=5.5$ MeV. The equation to calculate the optimal angle
$\theta$ (\ref{27}) reads as
\begin{equation}
\label{29}
(p^2+m^2)~\sin\theta-M\left(p\cos\theta+m\sin\theta\right)=0~,
\end{equation}
where
\begin{equation}
\label{30}
M=2G~\int^\Lambda \frac{d{\bf p}}{(2\pi)^3}\frac{p}{|p_4|}~\left(\sin\theta-
\frac{m}{p}\cos\theta\right)~.
\end{equation}
The constant of four-fermion interaction is $G=\frac{4}{2N_c}\widetilde G$ while
expressed in the same units as the mean energy functional in Eq.(\ref{26}). For
the NJL model Eq.(\ref{29}) makes it possible to contract a functional space
in which the minimum of mean energy functional can be
realized. This equation ~parameterizes the function $\theta(p)$ on the whole
interval $p\in[0,\Lambda]$ of searching the solution. Moreover, Eq.(\ref{29})
itself does not impose any restrictions on the parameter $M$ which may be any
real number. Then the functional (\ref{28}) simply becomes the function of
parameter $W(M)$. Now if one expresses the trigonometrical functions via
parameter $M$ it is possible to make the representation of minimizing function and the
result of its integration (\ref{30}) agree. As a result we receive three extremal
points, two of them correspond the minimal points with negative and positive
values of $M$ and the negative value conforms to the state of more stability.
The point of unstable equilibrium is located in the vicinity of coordinate  origin
$\sim m$. The induced quark mass for the parameter magnitudes fixed is $M=-335$ MeV
and the quark condensate
\begin{equation}
\label{37}
\langle\sigma|\bar q q|\sigma\rangle=\frac{i~N_c}{\pi^2}~\int_0^\Lambda
dp~\frac{p^2}{|p_4|}~(p\sin\theta-m\cos\theta)~,
\end{equation}
develops the magnitude of $\langle\sigma|\bar q q|\sigma\rangle=-
i~(245$ MeV$)^3$. The characteristic constant of the four-fermion interaction is
equal to $G/(2\pi^2)=1.34$. In what follows we rely on these quantitative results.

The situation, if the correlator $J_{\mu\nu}(p)$ is not equal to zero and has
the same form of the $\delta$-function in coordinate space, can be similarly
analyzed. The numerical analysis done teaches that its influence can be
essential but we do not show these results due to the lack of any phenomenological
estimates of the correlation function magnitude. The non-local version of the NJL
model in which the correlator has the separable form $I(p,q)=K(p)~K(q)$ can be
similarly analyzed. In fact, it again displays the above mentioned property which
replaces the functional analysis for the analysis of function dependence on some
parameter. Although one important difference does exist and it shows that the procedure
of integral cutting off is unnecessary for the functions $K(p)$. The regularization
is naturally performed by the $K(p)$ kernel and so strong regularization is caused by
the separable form interaction kernel. Certainly, such a property can manifest itself
in much weaker form for more realistic correlators.

\section{The Keldysh model}
Here we are going to analyse the limit in which the correlation function has a
$\delta$-function form in the momentum space
$$I({\bf p})=(2\pi)^3~G~\delta({\bf p})~.$$
This limit is an analogue of the Keldysh model which is well known in the
physics of condensed matter \cite{6} and the mean energy functional (\ref{26})
develops the following form in this case{\footnote{For sake of simplicity we do
not consider the contribution of correlator $J_{\mu\nu}(p)$.}}
\begin{widetext}
\begin{equation}
\label{31}
W(m)=\int \frac{d{\bf p}}{(2\pi)^3}~\left[|p_4|~\left(1-\cos\theta\right)-
G~\frac{p^2}{|p_4|^2}\left(\sin\theta-\frac{m}{p}\cos\theta\right)^2\right]~.
\end{equation}
\end{widetext}
The optimal values of angle $\theta$ are determined by the solutions of the
following equation
\begin{equation}
\label{32}
|p_4|^3~\sin\theta-2G~\left(p\cos\theta+m\sin\theta\right)
\left(p\sin\theta-m\cos\theta\right)=0~
\end{equation}
and we start analyzing these solutions in the chiral limit $m=0$. One of the
solutions corresponds to the zero angle $\theta=0$ but the non-trivial one takes
the form
\begin{equation}
\label{33}
\cos\theta=\frac{p}{2G}~.
\end{equation}
Both the positive and negative angles $\theta$ are suitable as the solutions
because of the parity (positive) property of the functional (\ref{31}) and these
(real) solutions (additional to the trivial one) exist on the limited momentum
interval $p<2G$. There are one real solution for the trivial angle and two imaginary
(complex-conjugate) solutions beyond this interval. Analyzing the NJL model above we
noticed its very convenient property when the solution $\theta(p)$ is defined on the
whole interval and, in fact, the functional is parametrized by a single number which
is the integral $M$.
\begin{figure}
\includegraphics[width=0.3\textwidth]{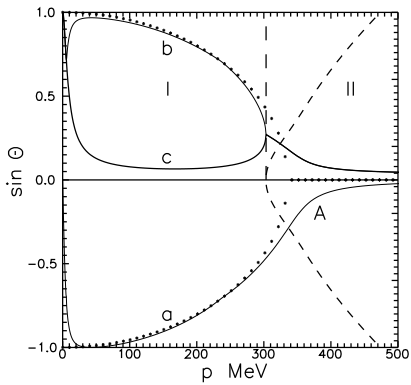}
\caption{Phase portrait of the Keldysh model, $\sin\theta$ as a function of
momentum $p$(MeV). The dotted curves correspond to the solutions in the
chiral limit $m=0$.}
\label{f1}
\end{figure}
In the Keldysh model the situation is much more sophisticated and
the phase portrait of its solutions in the chiral limit (for example,
$\sin\theta$ as a function of momentum $p$) consists of two arches
(with positive and negative $\sin\theta$, see Fig.1) and straight line
corresponding to the trivial solution. Thus, the semi-axis $p\in[0,\infty)$ can
be divided into two parts. There are three branches (solutions) at the interval
$p\in[0,2G]$, two of those correspond to the positive and negative angles
$\theta=\pm\arccos(p/2G)$ and the trivial one where $\theta=0$. At the interval
$p \in[2G,\infty)$ only one trivial solution $\theta=0$ exists and in order to
construct the solution on the whole semi-axis $p\in [0,\infty)$ one has to add
the trivial solution on the interval $p \in[2G,\infty)$ to any detached branch of
solutions on the interval $p\in[0,2G]$. It is easy to see that making use the imaginary
branches of solutions leads to the significant growth of energy and just because of
this fact they are uninteresting. The other potentially interesting functions $\theta(p)$
for which it is reasonable to search the functional minima could be received if the
interval $p\in[0,2G]$ is subdivided into smaller intervals and then for each
interval when continuing the function to the next interval (for example, to the direction
of the momentum $p$ increasing) to use two other branches as well as of the results of
continuation on the same branch. Apparently, it results in the piecewise continuous
function and unlike the NJL model here we have no parameter which restricts the
function and watches its integral characteristics. In the chiral limit all the solutions
(trajectories) constructed in such a way will acquire strictly fixed (finite) value of
the functional $W(0)$ (it will be observed that the functional does not contain the
derivatives of angle in momentum). For example, the trajectory which is going along
the top arch at the interval $p\in[0,2G]$ and continuing longer as a trivial solution
to the whole semi-axis leads to the magnitude
$$W_\pm(0)=-\frac{G^4}{15\pi^2}~,$$
(similarly for the top negative arch). The chiral condensate (\ref{37}) turns
out then to be
$$\langle\sigma|\bar q q|\sigma\rangle(0)=\frac{i~N_c~G^3}{2\pi}~,$$
(and for the solution along the negative arch we have the opposite sign). The
mean energy and chiral condensate equal to zero for the trivial solution, i.e.
($W_0(0)=0$, $\langle\sigma|\bar q q|\sigma\rangle_0(0)=0$).
Clearly, these piecewise continuous functions will lead to the magnitudes of
functional $W(0)$ which fill up the interval $[W_0(0),W_\pm(0)]$ densely, a
similar pattern takes place for the chiral condensate. With the natural parametrization
\begin{equation}
\label{theta}
\sin\theta=\frac{M_\theta}{(p^2+M_\theta)^{1/2}}~,
\end{equation}
we obtain for the mass $M_\theta$ which characterize the angle at the top arch
the following result
\begin{equation}
\label{34}
M_\theta=\left(4G^2-p^2\right)^{1/2}~.
\end{equation}
It is interesting to notice that then the respective energy of nontrivial
solutions $E(p)=\sqrt{p^2+M_\theta^2}$ becomes constant $E(p)=2G$.

After having done the analysis in the chiral limit which is shown by the dotted
lines in Fig. \ref{f1} we would like to comment on the situation beyond this
limit, i.e. where $m\neq 0$. The evolution of corresponding branches is available on the
same plot \ref{f1} where the behaviour of $\theta(p)$ as the function of momentum $p$
in MeV is shown for the solution of Eq.(\ref{32}). The semi-axis $p \in[0,\infty)$
where we are searching the solution can be subdivided into two sectors which are
demonstrated by the vertical dashed line on the plot. Three solutions denoted by $a$,
$b$ and $c$ are developing at the first sector denoted in Fig. \ref{f1} by I. Besides,
there are three solutions at the second sector denoted by II, one real solution
designated  as $A$ for the negative pairing angle and two complex-conjugate roots with
the positive  real parts. The imaginary parts of solutions are plotted in Fig. \ref{f1}
by the dashed lines. The solution $A$ in the domain II develops the behaviour of
$\theta\sim-\frac{2Gm}{p^2}$ with increasing momentum. As in the chiral limit the minimum
of the mean energy functional $W(m)$ can be obtained with the piecewise continuous
functions which are properly represented by the trajectories $aA$, $bA$, $cA$ (for real
solutions). The first symbol of this complicated designation implies the
branches $a$, $b$, or $c$ at the first sector, the second symbol corresponds to the branch
at the sector II. Thus, at low momenta we start with the solution of branches $a$, $b$
or $c$, then relevant solution passes to the branches interchanging its position in any
subinterval. But in any case there is only one way to continue the real solution
when momenta goes to infinity and it is related with the branch $A$ when the angle is
going to the zero value. Moreover, if the angle $\theta$ could take strictly zero value
in the sector II then the second term of Eq.(\ref{31}) leads to the singular contribution
coming from the term $\frac{m^2}{p^2}\cos^2\theta$ with the linear divergency  at large
momentum. Besides, the other terms develop the logarithmic divergencies as well. It is an
amusing fact the mean energy functional out of chiral limit goes to an
infinity at any nonzero value of current quark mass $m$ although in the chiral
limit $W(0)$ is well defined. (It is worthwhile to remember here the current  mass
singularity of zero mode approximation which was discovered in Ref. \cite{Ker}). The same
conclusion is valid for the chiral condensate (see Eq.(\ref{37}) in which the first and
second terms are developing the linear and quadratic divergencies, respectively. We could
conclude here that if the cut-off factor is not used in the integrals when dealing with
the solutions on the whole axis the functional $W(m)$ and quark condensate
$\langle\sigma|\bar q q|\sigma\rangle(m)$ are ill-defined.

Let us remember here that by definition the approximation (\ref{9}) should
describe the quark behaviour in the background of stochastic gluon field (which
is averaged) at low energies. Then it looks quite natural to introduce an
effective cut-off (in momentum) parameter $\widetilde \Lambda$. The condition for
factorization of gluon and quark field contributions gets broken at the momenta
above $\widetilde \Lambda$. In such a situation the dependence of mean energy
and quark condensate on the current quark mass is defined not only by the form
of integrand but by the value of parameter $\widetilde\Lambda$ as well. And if this
value is pretty large $\widetilde \Lambda \gg M$ (where $M$ is the dynamical quark mass)
the dependence on $m$ of all the observables is mainly defined by the magnitude of
cut-off parameter $\widetilde \Lambda$ because of the singular character of integrals
(for example, for the NJL model this magnitude could be estimated as $\sim 1 GeV$).
Obviously, it means in order to get the dependence of observables on the current quark
mass we need to draw essential additional information.

\begin{figure}
\includegraphics[width=0.3\textwidth]{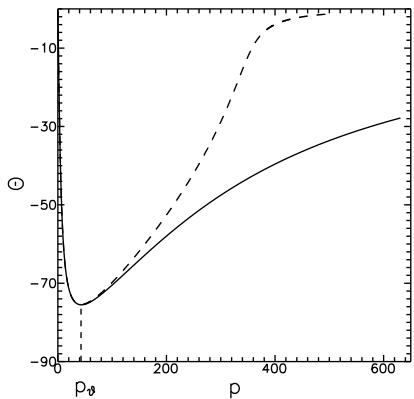}
 \caption{The equilibrium angle $\theta$ (in degrees) as a function of momentum
$p$ in MeV. The solid line shows the result of the NJL model and the dashed
line corresponds to the most stable branch of the Keldysh model, the current quark
mass is taken as $m=5.5$ MeV.}
  \label{f4}
\end{figure}

As to the possible interpretation of the singularities available in the mean
energy functional we could assume, for example, the mechanism similar to the
Cooper pairing which takes place at every scale of the increasing momenta
$\widetilde \Lambda$, $\widetilde \Lambda_1$, \dots. Certainly, we should correct
the existing results about four-quark interaction potential to put the pairing
effect on realistic ground. For example, the contribution of the stochastic
configurations like the small size instantons which is exponentially suppressed
is hardly relevant to provide an efficient pairing mechanism for the momenta above
$\widetilde \Lambda$. Apparently, the hard gluon exchange looks like a more adequate
mechanism at small distances. Then the gluon correlation function in
Eq.(\ref{9}) should be transformed in the corresponding gluon propagator.
The effective four-quark interaction we are interested in can be derived by the
quasi-average formalism \cite{3} which approximates smoothly the infrared and
~ultraviolet momentum regions although an alternative scenario could also be
quite meaningful (see, for instance, the discussion in \cite{9}). The fact that
the Cooper attraction is still large enough despite the coupling constant weakening
could signal the dominance of more fundamental fields at very small distances.
(Here it is easy to see all the different models might be classified by the
convergence of the integral over momentum with the constant of four-fermion
interaction
$$I_G=\int dp~G(p)~,$$
as the integrand. The model falls under the category of a singular one if this
integral diverges.)

In Fig. \ref{f4} we compare the equilibrium angles $\theta$ for the NJL model
(solid line) and the $aA$ solution of the Keldysh model (dashed curve) as the
functions of momentum $p$ in MeV with the current quark mass $m=5.5$ MeV.
It is interesting to notice that out of the chiral limit the solution
(which has a spherical symmetry) passed over zero at $p=0$ (see, Eq.(\ref{35})).
Besides, Fig.\ref{f4} demonstrates that out of the chiral limit the pairing
process becomes essential not at zero momentum value (as it takes place
in the chiral limit) but it is shifted to the magnitude about
$p_{\theta}\sim 40$ MeV  for the fitting parameters used. For example in the NJL
model it can be obtained
$$p_\theta=\left[m~|M-m|\right]^{1/2}~.$$
The quantity $r_{\theta}=1/p_{\theta}$ determines the characteristic size of the
region which is efficient for the pairing process. In the chiral limit this
region is formally extending to infinity. The curves shown in the plot correspond
to the opposite, in a sense, limiting regimes and it is interesting to evaluate
where the model with more realistic correlator could be found out.

One of the important motivations to study the Keldysh model was the question of
a natural regularization which presents for the interaction with separable
kernel. We have seen that in the chiral limit for the kernel with most extensively
expressed regularizing property as, for example, the momentum $\delta$-function,
both the mean energy and chiral condensate are well defined. Out of the chiral
limit the unexpected singularity appears. In Ref. \cite{weAr} it was discussed
the possibility of continuing the mean energy functional and the quark condensate
by performing the respective regularization. As well known the meson masses in the
NJL model can be presented by the quark condensate what hints the corresponding
expressions in the Keldysh model could be singular as well and one needs to perform
another regularization to provide them with clear physical meaning. However, despite
the present singularity of chiral condensate the meson observables are finite and
are well matched with the experimental mass scale (see \cite{MVZ}).
The reason to have these meson observables as the smooth functions of current
quark mass is in the regularizing role of additional vertex formfactors which enter
the meson mass formulae. Then we may summarize that it does not make sense to debate
about an absolute value of quark condensate (in vacuum) for considered mechanism
of spontaneous chiral symmetry breakdown because its magnitude depends on the
particular observable (characteristic momentum of saturation) which is used for
extracting this data. In the Appendix II we compare the results obtained in the
Hartree-Fock-Bogolyubov approach for the NJL and Keldysh models with the results
of mean field approximation.
\begin{figure}
\includegraphics[width=0.3\textwidth]{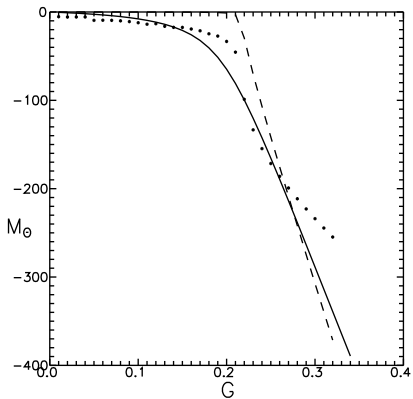}
 \caption{The parameter $M_\theta$ (solid and dashed lines) and quark condensate
(without an imaginary unit, in power $1/3$) in MeV (shown by points calculated
in the chiral limit $m=0$) as a function of the constant $G$ for the Gaussian
correlator. The solid line is calculated with the current
quark mass $m=5.5$ MeV and the dashed line is calculated in the chiral limit.}
  \label{f6}
\end{figure}

\section{The exponential and Gaussian correlators}
Here we turn to a more realistic situation and analyze the solutions possessing
a spherical symmetry in the regime where the correlation function $I({\bf x})$ is
rather quickly decreasing with the distance increasing. Performing the
integration over the azimuthal angles we can get the equation to derive the optimal
angle in the following form
\begin{widetext}
\begin{equation}
\label{35}
|p_4|^2\sin\theta-\frac{4G}{\pi}\left(\cos\theta+\frac{m}{p}\sin\theta\right)
\int\!\!\!\!\int_0^\infty dq dx~\frac{q}{|q_4|}\left(q\sin\theta'-
m\cos\theta'\right)~I(x)~\sin(px)\sin(qx)=0.
\end{equation}
\end{widetext}
Considering the solution behaviour at high momentum values $p$ we are interested
in analyzing solutions in which the angle $\theta$ is going to zero magnitude.
Assuming the $\theta$ value to be rather small we expand (\ref{35}) up to the
terms of the $\theta$ order and have
$$p^2\theta-\frac{4G}{\pi}\int_0^\infty dq(q~\theta'-m)~I(p,q)=0~.$$

If the function $\theta$ is decreasing faster than $1/q$ the most essential
contribution to the integral comes from the term proportional to $m$ and if the
kernel $I(p,q)$ is integrable the asymptotic behaviour has the following form
$$\theta=-\frac{4G~m}{\pi~p^2}~\int_0^\infty dq~I(p,q)~.$$
Let us consider now two concrete examples, with exponential behaviour of the
correlator $I({\bf x})=\exp{(-a~|{\bf x}|)}$, and with the Gaussian behaviour
$I({\bf x})=\exp{(-a^2~{\bf x}^2)}$. The integration over $x$ can be
performed exactly for both cases and the kernels of integral equations look like
\begin{eqnarray}
&&\int_0^\infty dx~ e^{-ax}~\sin(px)\sin(qx)=\nonumber\\
&&=\frac{a}{2}~
\left(\frac{1}{a^2+(p-q)^2}-\frac{1}{a^2+(p+q)^2}\right)~,\nonumber
\end{eqnarray}
for the exponential correlator and as
\begin{eqnarray}
&&\int_0^\infty dx~ e^{-a^2x^2}~\sin(px)\sin(qx)=\nonumber\\
&&=\frac{\sqrt{\pi}}{4a}~
\left(e^{-\frac{(p-q)^2}{4a^2}}-e^{-\frac{(p+q)^2}{4a^2}}\right)~,\nonumber
\end{eqnarray}
for the Gaussian one. Now let us hold the contribution of the first term only at
large momentum values $p$ for both examples. Then as a result the corresponding
asymptotic behaviours are expressed by the constants which are defined by the
integrals with the kernels $I(p,q)$. It allows us to conclude that we have again the
singular functional for the mean energy out of the chiral limit.
The parameter $M_\theta$ (see Eq.(\ref{theta}) and the quark condensate as
functions of the constant $G$ for the Gaussian correlator (both obtained by the
numerical computation of Eq.(\ref{35})) are depicted in Fig. \ref{f6}.
The solid line demonstrates the solution with the current quark mass
$m=5.5$ MeV and the dashed line is calculated in the chiral limit as the quark
condensate presented by the points. The intrinsic change of the parameter
$M_\theta$ generation out of the chiral limit is easily seen. The similar
features are observed for the exponential correlator as well.

Unfortunately, it is a very serious problem to get all the solutions of the
nonlinear integral equation (\ref{35}) and here we are working with only one of
its (the most stable) branches. As it was demonstrated above such a situation
generates a lot of difficulties for extracting a reliable information
on the observables out of the chiral limit. Due to this reason we calculate here
the dynamical quark mass ($M=-335$ MeV) and chiral condensate
($|\langle\sigma|\bar q q|\sigma\rangle|=(245$ MeV$)^3$)
in the chiral limit collating the dynamical quark mass with $M_\theta$ and
fitting the parameters $a$ and $G$. The parameter $a$ for the
exponential and Gaussian correlators reads as
$$a_{ex}=0.15~{\mbox{GeV}},~~~a_{gs}=0.16~{\mbox{GeV}}~.$$
The most suitable values of $G$ are equal to
$$G_{ex}=0.35~,~~~G_{gs}=0.31~$$
and $|M_\theta^{ex}|=338$MeV, $|\langle\sigma|\bar q
q|\sigma\rangle_\theta^{ex}|=(228$ MeV$)^3$,
$|M_\theta^{gs}|=340$MeV, $|\langle\sigma|\bar q
q|\sigma\rangle_\theta^{gs}|=(245$ MeV$)^3$.
Actually we can collate the gotten value of the four-fermion interaction
constant with the packing fraction parameter of which the basic one for
the instanton vacuum model is $\widetilde G=(4\pi^2)^2~n\rho^4$.
The result of this exercise $n\rho^4\sim 10^{-3}$ is quite
realistic. However, there is a pretty serious discrepancy in the estimates of
characteristic configuration size (we should keep in mind the calculations are
done in the chiral limit).
\begin{figure}
\includegraphics[width=0.3\textwidth]{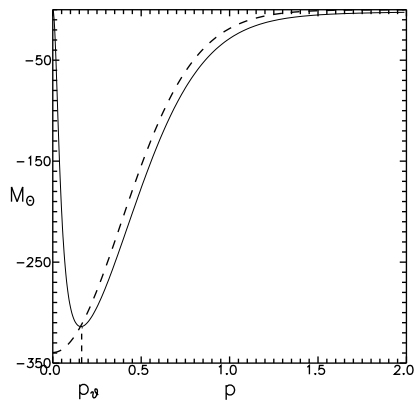}
 \caption{The parameter $M_\theta$ in MeV as a function of momentum $p$ in GeV
which corresponds to the best fit of 'experimental data'. The solid curve is
calculated for the Gaussian correlator with $m=5.5$ MeV, the dashed line is
calculated for the same
correlator but in the chiral limit.}
  \label{f7}
\end{figure}

The parameter $M_\theta$ as a function of momentum $p$ in GeV calculated with
the parameters corresponding the best fit of 'experimental data' is
depicted in Fig. \ref{f7}. The solid curve is obtained for the Gaussian
correlator with $m=5.5$ MeV and the dashed line is calculated for the same
correlator but in the chiral limit.
Here we do not mention the results obtained for the behaving exponentially  correlator because
they practically coincide with the results for the Gaussian correlator.
The parameter $p_{\theta}$ is estimated at the current quark mass $m=5.5$ MeV to
be as $p_{\theta}\sim 150$ MeV, i.e. of the $\pi$-meson mass order. The
treatment of the correlator with instanton profile together with the detailed
analysis of exponential and Gaussian correlators is worthy of special paper and
will be studied in the next paper.

\section{\bf Conclusion}
In the present paper we undertake the efforts to systematically study the
quark's behaviour in various ensembles of stochastic gluon fields developing
simple ensemble approximation which is grounded on the circumstantial analysis
of two-particle correlation function. An approximate procedure developed enables to
calculate the effective functional for the mean energy and to estimate the ground
state parameters within the Hartree-Fock-Bogolyubov approach. The models with the
exponential and Gaussian behaviours of correlators are
analyzed in the chiral limit and their parameters are fitted. The results
obtained are used to estimate the characteristic region size $r_\theta$ in which
the possible processes of quark--anti-quark pairing might become significant.
This size was estimated to be $r_\theta\sim 1/40$ MeV$^{-1}$ for
the parameters inherent in the NJL model. For the exponential and Gaussian
correlators this estimate looks like $r_\theta\sim 1/150$ MeV$^{-1}$.

Besides, we clearly demonstrate the presence of singularity in the mean energy
functional outside the chiral limit. Finally, let us emphasize the quark
ensemble characteristics discussed in the paper are not physically
observable and in order to make the intelligent conclusion about the
model effectiveness one should explore, for example, the meson correlation
function. In fact, it has been done for the Keldysh model in \cite{MVZ}
and result turned out to be quite encouraging. Despite the singular
 character of the mean energy of the system the meson observables are finite,
quite recognizable and comparable with the energy scale coming from an experiment.

The authors are very thankful to I.V. Anikin, B.A. Arbuzov, A.E. Dorokhov, S.B.
Gerasimov, E.-M. Ilgenfritz, N.I. Kochelev, S.N. Nedelko, A.E. Radzhabov, A.M.
Snigirev, O.V. Teryaev and M.K. Volkov for numerous fruitful discussions and constructive
criticism. It is a pleasure to thank the referee for instructive comments on an earlier
version of this paper. This work was supported by the INTAS Grant 04-84-398 and the NAS of
Ukraine project "Fundamental features of physical systems under extreme conditions".

\section{Appendix I}
Here dealing with the simple quantum mechanical example we demonstrate the
difference between the description based on the averaging of density matrix and
the approach in which the averaging of the generating functional is used
(considering the model system in a real time).

1. Let us suppose the particle described by the stationary Hamiltonian $H_0$ is
also affected by the time dependent force $f(t)$. Then the particle state is
circumscribed by the Schr\"odinger equation
$$i \dot \Psi= H~\Psi~,~~H=H_0+V~,~~V=f~x~,$$
where $\Psi$ is the corresponding wave function and we search the perturbative
solution as
$$\Psi=\Psi^{(0)}+\Psi^{(1)}+\dots$$
We expand the wave function of the zeroth order in the eigenfunctions of
Hamiltonian $H_0$ as
$$\Psi^{(0)}=c_n~e^{-i\lambda_n~t}\psi_n~,~~H_0~\psi_n=\lambda_n~\psi_n~,$$
the constants $c_n$ are defined by the initial condition here, and the next
perturbative orders are calculated in the following form
$$\Psi^{(j)}=d^{(i)}_n~e^{-i\lambda_n~t}\psi_n~,$$
where the coefficient $d^{(i)}_n$ is determined by the iterations as
$$d^{(j)}_n(t)=-i~(\stackrel{*}\psi_n \!x \psi_m)\int_0^{t}
d\tau ~f(\tau)~ e^{i(\lambda_n-\lambda_m)\tau}~d^{(j-1)}_m(\tau),$$
and $(\stackrel{*}\psi_n\! x \psi_m)$ stands here for the matrix element over
the eigenfunctions of $H_0$, $j=1,2,\dots$, $d^{(0)}_n=c_n$.
The energy operator after averaging over the final state $\Psi$ can be presented
in the form of trace $\mbox{ Tr}~\{H\rho\}$ with the pseudo-density matrix
$\rho=\Psi \stackrel{*}\Psi$. In general, an energy being averaged over such a
pseudo-matrix will be time dependent but at analysing the quasi-stationary
processes, for example, it might be useful to study its averages in time as reads
$$\overline{\mbox{ Tr}~\{H\rho\}}=\int_0^{T} d t~ \mbox{ Tr}~\{H\rho\}/T~.$$
For the sake of clarity we suppose for the force mean value that $\bar f=0$.
Then a nontrivial contribution into the interaction mean energy comes from the
cross terms of the zeroth and first orders of perturbation expansion
$\stackrel{*}{\Psi^{(0)}}$ $\Psi^{(1)}$,
$\Psi^{(0)}$ $\stackrel{*}{\Psi^{(1)}}$ and we have
\begin{eqnarray}
\label{ap1}
\mbox{ Tr}~\{V\rho\}&=&i~ c_k \stackrel{*}c_m (\psi_k x\! \stackrel{*}\psi_n)
(\psi_n x \!\stackrel{*}\psi_m)\times
\nonumber\\ [-.2cm]
\\ [-.25cm]
&\times&e^{i(\lambda_n-\lambda_k)t}\int_0^{t}
d\tau~ f(\tau)f(t)~e^{i(\lambda_k-\lambda_n)\tau}+\mbox{c.c}~.\nonumber
\end{eqnarray}

At estimating the impact of the stochastic force its contribution may be
factorized with a help of the corresponding correlation function
$\overline{f(\tau) f(t)}\sim \overline{f^2}~F(\tau-t)$
if the characteristic frequencies $\lambda_n$ are smaller then the stochastic
ones. In the particular case of the 'white noise' (when the profile function $F$ has
the $\delta$-function shape) the time dependence in the intermediate states
$\stackrel{*}\psi_n$,  $\psi_n$ in Eq. (\ref{ap1}) disappears (see the
corresponding exponentials depending on $\lambda_n$).
Due to the assumed completeness of eigenvalues basis of $H_0$, i.e.
$\sum_{n}|\psi_n\rangle \langle \stackrel{*}\psi_n\!\!|=1$,
Eq. (\ref{ap1}) may be presented as
$$\mbox{ Tr}~\{V\rho^{(2)}\}\simeq i c_k \stackrel{*}c_m (\psi_k x^2
\stackrel{*}\psi_m)\overline{f^2}~\overline{e^{i(\lambda_m-
\lambda_k)t}}+\mbox{c.c}~.$$
It allows us to conclude that the resulting averaged final state density matrix
is weighed with the effective 'potential' of form $\overline{f^2}~x^2$. The
similar results can be received in the next perturbative orders. The cluster decomposition
of stochastic exponential $e^{i~fx}$ is practical to demonstrate that the same
results for the effective 'potential' of interaction take place for a 'white noise'
at averaging the generating functional $\langle \Psi\rangle$ (as it is claimed in
the first section of this paper). In general consideration there appears a certain
nonlocal 'potential' and its properties are dependent of the system state.

2. Now let us turn to the description in terms of a density matrix only. It is
defined by the following equations
\begin{equation}
\label{ap4}
i~\dot \rho=H~\rho-\rho~H'~,
\end{equation}
and the density matrix is dependent on the coordinates and time $\rho(x,y;t)$.
The operator $H$ is acting on the coordinate $x$ and the operator $H'$ is acting
on the coordinate $y$. In the zeroth order of perturbative expansion we have
$$\rho^{(0)}=c_{nm}~e^{i\lambda_n t} \stackrel{*}\psi_n(y)~ e^{-i\lambda_m t}\psi_m(x)~,$$
where $c_{nm}$ is a hermitian matrix which is defined by the initial data. In
the first order of the perturbation series we present the solution in the
following form
$$\rho^{(1)}=d^{(1)}_{nm}~e^{i\lambda_n t} \stackrel{*}\psi_n(y)~ e^{-i\lambda_m
t}\psi_m(x)~.$$
It is possible to have for the matrix $d^{(1)}$ the representation as
\begin{eqnarray}
d^{(1)}_{nm}=&-i& c_{nk}~ (\stackrel{*}\psi_m x \psi_k)~\int_0^{t}
d\tau ~f(\tau)~e^{i(\lambda_n-\lambda_k)\tau}+\nonumber\\
&+i& c_{km}~ (\stackrel{*}\psi_k y \psi_n)~
\int_0^{t} d\tau ~f(\tau)~e^{i(\lambda_k-\lambda_m)\tau}~.\nonumber
\end{eqnarray}
Then for the density matrix $\rho^{(1)}$ the following form appears (in order to
get it we have to interchange the indices $m$ and $n$ in the second term)
\begin{widetext}
\begin{eqnarray}
\rho^{(1)}=&-&i c_{nk}~ (\stackrel{*}\psi_m x \psi_k)~e^{i(\lambda_n-\lambda_m)t}~
\stackrel{*}\psi_{n}(y)~\psi_m(x)~\int_0^{t} d\tau~ f(\tau)~
e^{i(\lambda_n-\lambda_k)\tau}+\nonumber\\
&+&i c_{kn} ~(\stackrel{*}\psi_k y \psi_m)~e^{i(\lambda_m-\lambda_n)t}~
\stackrel{*}\psi_{m}(y)~\psi_n(x)\int_0^{t} d\tau~ f(\tau)~
e^{i(\lambda_k-\lambda_n)\tau}~.\nonumber
\end{eqnarray}
\end{widetext}
Due to the hermitian property of density matrix we have $c_{kn}=\stackrel{*}c_{nk}$
the second term is complex conjugate with the first one at the coinciding arguments
$x=y$. Then calculating the mean interaction energy
$\mbox{Tr}\{ V\rho^{(1)}(x,y;t)|_{y\to x}\}$,
we are convinced that the result is identical to what we found out at the
beginning of this section.

Let us consider the solution for density matrix in the operator form (and
without specifying the basis functions) as
$$\rho=e^{i(H'_0-H_0)t}~\widetilde \rho~,$$
where the matrix $\widetilde \rho$ is determined by the solution of following
integral equation
$$\widetilde \rho(t)=-i \int_0^{t} \!\!d\tau e^{i(H_0-H'_0)\tau}  f(\tau)(x-
y)e^{i(H'_0-H_0)\tau}\widetilde \rho(\tau)+\widetilde \rho(0).$$
and effective interaction Hamiltonian is given by the operator expression as
\begin{eqnarray}
&&\mbox{Tr}\{H\rho\}=\mbox{Tr}\left\{[H_0+f(t)~ x]~(-i)~
e^{i(H'_0-H_0)t}\times\right.\nonumber\\
&&\left.\times\int_0^{t} d\tau~ e^{i(H_0-H'_0)\tau} ~f(\tau)~(x-y)~
\rho(x,y;\tau)|_{y\to x}\right\}.\nonumber
\end{eqnarray}

3. Further we analyse some details of the particular exercise which admits of
receiving the overt expressions and consider the forced oscillations defined by
the Hamiltonian
$$H_0=-\frac{1}{2m}\frac{d^2}{d x^2}+\frac{m\omega^2}{2}x^2~.$$
The ~continual integral is exactly calculated for this exercise \cite{1} and the
presentation of pseudo-density matrix which we are interested in looks like
\begin{eqnarray}
\label{ap1_1}
\stackrel{*}{\psi}(y_2,t_2)\psi(x_2,t_2)&=&\int\!\!\!\int_{-\infty}^{\infty}\!\!
dx_1dy_1~\stackrel{*}K(y_2,t_2;y_1,t_1)\times\nonumber\\[-.2cm]
\\ [-.25cm]
&\times&K(x_2,t_2;x_1,t_1)\stackrel{*}\varphi(y_1,t_1)\varphi(x_1,t_1)~,\nonumber
\end{eqnarray}
where $\varphi(x_1,t_1)$ is an initial state. The transformation kernel is
expressed by the overt formula like
\begin{equation}
\label{ap1_2}
K(x_2,t_2;x_1,t_1)=\left(\frac{m\omega}{2\pi i \sin \omega T}\right)^{1/2}~e^{i S}~,
\end{equation}
where the action is given as
$$S=\frac{m\omega}{2 \sin \omega T}[\cos \omega T (x_2^2+x_1^2)-2 x_2 x_1+2x_2
\phi_2+2x_1 \phi_1-F],$$
and the phase factor have the following form
\begin{eqnarray}
&&\phi_1=\frac{1}{m\omega}\int_{t_1}^{t_2} d\tau f(\tau)\sin \omega(t_2-
\tau)~,\nonumber\\
&&\phi_2=\frac{1}{m\omega}\int_{t_1}^{t_2} d\tau f(\tau)\sin \omega(\tau-
t_1)~,\nonumber
\end{eqnarray}
$T=t_2-t_1$. The term $F$ depends on the time ~parameters only and is immaterial
because it is cancelled in the exponential exponent of pseudo-density matrix. We
introduce the new variable $x_2=\widetilde x_2+\phi_1$ and transform the exponential
exponent in the $K$ kernel as
\begin{eqnarray}
&&\cos \omega T (x_2^2+x_1^2)-2 x_2 x_1+2x_2 \phi_2+2x_1 \phi_1=\nonumber\\
&&=\cos \omega T (\widetilde x_2^2+x_1^2)-2 \widetilde x_2 x_1+
2(\cos \omega T \phi_1+\phi_2)\widetilde x_2+\nonumber\\
&& +\cos \omega T \phi_1^2+2\phi_1\phi_2~.\nonumber
\end{eqnarray}
The similar transformations should be done in the kernel $\stackrel{*}K$ with
the variable $y_2=\widetilde y_2+\phi_1$.
The formulae take more convenient form if we introduce the auxiliary factor
$\bar\phi_1=(\cos \omega T \phi_1+\phi_2)/sin~\omega T$
which can be written down by the simple transformations in the following form
$$\bar\phi_1=\frac{1}{m\omega}\int_{t_1}^{t_2} d\tau f(\tau)\cos \omega(t_2-\tau)~.$$
Making use the well known representation of the kernel $K$ for non-perturbated
oscillator ($f=0$), see \cite{1},
$$K_0(x_2,t_2;x_1,t_1)=\sum_{n=0}^{\infty} e^{-i\lambda_n
T}~\stackrel{*}\psi_n(x_2)\psi_n(x_1).$$
It is easy to understand that the important terms of pseudo-density matrix
kernel $\stackrel{*}K \stackrel{}K$ at $y_2\to x_2$ are represented in the similar form
\begin{eqnarray}
\label{ap1_3}
\stackrel{*}K \stackrel{}K&=&
\sum_{n=0}^{\infty} e^{i\lambda_n T}~\psi_n(\widetilde y_2)\stackrel{*}\psi_n(y_1)~
e^{-i m\omega\bar\phi_1\widetilde y_2}\times\nonumber\\
&\times& e^{i m\omega\bar\phi_1\widetilde x_2}
\sum_{m=0}^{\infty} e^{-i\lambda_m T}~\stackrel{*}\psi_m(x_2)\psi_m(x_1)~|_{y_2=x_2}.\nonumber
\end{eqnarray}
The following matrix element
$$\int_{-\infty}^\infty d x_2~ \psi_n(\widetilde y_2)~e^{-i
m\omega\bar\phi_1\widetilde y_2}~ H(x_2)~
e^{i m\omega\bar\phi_1\widetilde x_2}\stackrel{*}\psi_m(x_2)|_{y_2=x_2}~$$
will be faced at calculating the mean energy.
Now moving the exponential $e^{im\omega\bar\phi_1\widetilde x_2}$ to the left
and changing the variable $x_2=\widetilde x_2+\phi_1$ in the Hamiltonian
$H(x_2)$ we obtain the representation
\begin{eqnarray}
\label{ap1_heff}
&&e^{-i m\omega\bar\phi_1\widetilde y_2}~H(x_2)~e^{i m\omega\bar\phi_1\widetilde x_2}=
H_0(\widetilde x_2)-\omega \bar\phi_1~ i\frac{d}{d \widetilde x_2}+\nonumber\\
[-.2cm]
\\ [-.25cm]
&&~~~~~~~~~~~~~+(m\omega^2\phi_1+f)\widetilde x_2
+\frac{m\omega^2}{2}(\phi_1^2+\bar\phi_1^2)+f\phi_1.\nonumber
\end{eqnarray}
which allows us to see that the mean energy calculated over the final state is
expressed by the diagonal elements and matrix elements of coordinate and
momentum as well
\begin{eqnarray}
&&\int_{-\infty}^{\infty}d x_2 \stackrel{*}K ~\stackrel{}H(x_2)
\stackrel{}K=\sum_{n=0}^\infty H_{n, n}~\psi_n(x_1)\stackrel{*}\psi_n(y_1)+\nonumber\\
&&~~~~~~~~~~~~~~~~~
+\sum_{n=0}^\infty H_{n, n-1}~e^{i\omega T}\psi_n(x_1)\stackrel{*}\psi_{n-
1}(y_1)+\nonumber\\
&&~~~~~~~~~~~~~~~~~
+\sum_{n=0}^\infty H_{n-1, n}~e^{-i\omega T}\psi_{n-
1}(x_1)\stackrel{*}\psi_{n}(y_1)~.\nonumber
\end{eqnarray}
where
\begin{eqnarray}
&&H_{n,n}=\left(n+\frac{1}{2}\right)\omega+\frac{m\omega^2}{2}(\phi_1^2+\bar\phi
_1^2)+f \phi_1~,\nonumber\\
&&\stackrel{}H_{n, n-1}=\stackrel{*}H_{n-1, n}=
\left[m\omega^2 (i\bar \phi_1+\phi_1)+f\right]\left(\frac{n}{2 m
\omega}\right)^{1/2}~.\nonumber
\end{eqnarray}

4. Now we would like to analyse the example of oscillations initiated by a
periodic perturbation defined as
$$f(t)=F\sin \Omega t~.$$
Then the phase factors develop the following form
\begin{eqnarray}
\label{ap1_ph}
&&m\omega^2\phi_1=F~\frac{\omega}{\Omega^2-\omega^2}
(\Omega\sin \omega T-\omega \sin \Omega T)~,\nonumber\\[-.2cm]
\\ [-.25cm]
&&m\omega^2\bar \phi_1=F~\frac{\Omega \omega}{\Omega^2-\omega^2}~
(\cos \Omega T-\cos \omega T)~,\nonumber
\end{eqnarray}
(for the sake of simplicity we take the parameter as $t_1=0$). As in the limit
of classical mechanics these expressions include a resonance behaviour at
coinciding the external frequency $\Omega$ and oscillator frequency $\omega$,
and in the resonance vicinity ($\Omega=\omega+\varepsilon$ with the small
deviation $\varepsilon$ from the oscillator frequency) the motion
behaves as the beats, i.e. the small oscillations with the frequency
$\omega$ and large amplitude. Now we are going to resolve the corresponding
classical equation
$$\ddot x_c +\omega^2 x_c=-f/m~,$$
with the initial conditions as $x_{c}(0)=0$, $\dot x_{c}(0)=0$
$$x_{c}=\frac{F}{m\omega}\frac{\Omega\sin \omega t-\omega
\sin \Omega t}{\Omega^2-\omega^2}~.$$
Comparing Eqs.(\ref{ap1_ph}) and $x_c$, $\dot x_c$ we are able to express the
phase factors $\phi_1$ and $\bar \phi_1$ as the classical coordinates
$\phi_1=x_{c}$ and velocity $\bar \phi_1=\dot x_{c}/\omega$. In particular,
the correction to the diagonal element of effective Hamiltonian (\ref{ap1_heff})
can be presented in the following form
$$\frac{m\omega^2}{2}(\phi_1^2+\bar\phi_1^2)+f\phi_1
=\frac{m}{2}\dot x_c^2+\frac{m \omega^2}{2} x_c^2+ f x_c~.$$

Averaging the mean energy with pseudo-density matrix we get the quadratic form as
\begin{equation}
\label{ap1_form}
\mbox{Tr}\{H\rho\}|_{y_2=x_2}=\sum_{n,m=0}^\infty
\stackrel{*}c_n~\stackrel{}H_{n,m}~\stackrel{}c_m~,
\end{equation}
with the coefficients $c_n$ defined by the initial state and normalized as
$\sum |{c_n}|^2=1$. In the considered situation of periodic force acting for
very long (unlimited) time it becomes clear the value of mean energy received is time
dependent (analogously to the classical description) and it means a certain asymptotic
value for mean energy (as for other observables)
simply does not exist\footnote{Apparently, this almost obvious fact was
underrated for rather long time.}. Physical meaning of this fact appears quite
transparent. The quantum system is carrying out the
repeated transitions to the upper levels of excited state and back (these
transitions are controlled by pseudo-density matrix) eventually resulting in some
quasi-stationary process which can be pithily
characterized by some observable values averaged in time. Thus, the averaged
magnitude of diagonal element of pseudo-density matrix
$\overline{H_{n,n}}=\int_0^{T} d t~H_{n,n}/T$ takes the form
\begin{widetext}
\begin{eqnarray}
\label{ap1_enfin}
\overline{H_{n,n}}=\left(n+\frac{1}{2}\right)\omega&+&
\frac{F^2}{m}\frac{\Omega^2+3\omega^2}{4(\Omega^2-\omega^2)^2}
+\frac{3 F^2}{8 m}\frac{1}{\Omega^2-\omega^2}\frac{\sin 2 \Omega T}{ \Omega
T}+\nonumber\\
&+&
\frac{F^2}{2 m\omega^2}\frac{\Omega\omega}{\Omega^2-\omega^2}
\left[\frac{\Omega-2\omega}{\Omega-\omega}\frac{\sin (\Omega-\omega) T}{(\Omega-
\omega)T}
-\frac{\Omega+2\omega}{\Omega+\omega}\frac{\sin (\Omega+\omega)
T}{(\Omega+\omega)T}\right]~.\nonumber
\end{eqnarray}
\end{widetext}
Then it is not difficult to see that asymptotically a quasi-stationary regime of
quantum ensemble oscillations as the whole can be set in, indeed, and now the
question of interest is to determine the minimum of functional (\ref{ap1_form}) which
corresponds to some effective ground state of the system while under the external
influence. The effective Hamiltonian (\ref{ap1_heff})
$$H_{eff}(x)=e^{-i m\omega\bar\phi_1\widetilde y}~ H(x)~
e^{i m\omega\bar\phi_1\widetilde x}~,$$
can be presented by using the classical variables $x_c$, $\dot x_c$
in the following form
$$H_{eff}(x)=\frac{\left(\hat p+p_c\right)^2}{2}+\frac{m\omega^2}{2}\left(\hat
x+x_c\right)^2+f(\hat x+x_c),$$
where $p_c=m \dot x_c$. This quantity (at the certain conditions) may be treated
in such a way that it is practical to search the ground state with the biased
coordinate $x_c$ and momentum $p_c$. Certainly, the treatment of excited states turns
out the nontrivial problem in this situation. At every time moment the pseudo-density
matrix is a pure magnitude because the equality $\rho^2=\rho$ is identically valid.
However, it is possible to estimate the purity degree of trail quasi-stationary state
$\varsigma=\mbox{Tr}~\{\overline{\rho^2}\}$,
with the time averaged density matrix and to find such states which allow us to
develop a description close to one in the terms of the Schr\"odinger equation.

The density matrix formalism is very practical in more general situations, for
example, at studying the influence of other quantum ensembles on a particle. It
is very actively discussed and developing (being often quite far from our concerns)
\cite{stoch} but our purpose here was to illustrate the difference in describing
a system with averaging a density matrix and averaging a generating functional.

\section{Appendix II}
The standard way to formulate an effective theory is to use the path integral
formalism. In order to transit to such a description we should construct the
corresponding Lagrangian action density from the effective Hamiltonian (\ref{9})
\begin{equation}
\label{36}
{\cal L}=\bar q (i \gamma_\mu \partial_\mu+im) q-G'\bar q~t^a\gamma_\mu q
\int d{\bf y} I_{\mu\nu}({\bf x}-{\bf y}) \bar q' t^a\gamma_\nu q',
\end{equation}
where $q=q({\bf x},t)$, $\bar q=\bar q({\bf x},t)$, $q'=q({\bf y},t)$, $\bar q'=\bar
q({\bf y},t)$. For the highest order in $N_c$ the sum of colour group generators looks
like $\sum^{N_c^{2}-1}_{a=1}t^a_{ij}t^a_{kl}\approx\frac{1}{2}~\delta_{il}\delta_{kj}$.
For the sake of simplicity we consider the correlator of the following form only
$I_{\mu\nu}({\bf x}-{\bf y})=\delta_{\mu\nu}~I({\bf x}-{\bf y})$.  Using the
Fierz transformation
$\gamma_\mu \bigotimes \gamma_\mu=1 \bigotimes 1 +i\gamma_5 \bigotimes i\gamma_5
-\frac{1}{2}\gamma_\mu \bigotimes \gamma_\mu -\frac{1}{2}\gamma_\mu\gamma_5
\bigotimes \gamma_\mu\gamma_5$, and holding only the scalar contribution we
receive in the mean field approximation the following effective Lagrangian density
\begin{equation}
\label{377}
{\cal L}=\bar q~(i \gamma_\mu \partial_\mu+im)~q-G'\int d{\bf y}~
I({\bf x}-{\bf y})~ \langle \bar q~q'\rangle~\bar q'~q~.
\end{equation}
The brackets in this expression imply the calculation of the corresponding
averages. The self-consistency condition of approximation which may be
formulated as the following integral equation
\begin{equation}
\label{38}
-iM({\bf p})=\int \frac{d q}{(2\pi)^4}~G'~I({\bf p}-{\bf q})~Tr~\frac{1_c}{\hat
q +im+iM({\bf q})}~,
\end{equation}
allows us to calculate the quark mass. Integrating over the fourth
component of momentum
\begin{eqnarray}
&&\int^{\infty}_{-\infty}\frac{d q_4}{2\pi}~\frac{1}{q_4^2+{\bf q}^2+(m+M({\bf
q}))^2}=\nonumber\\
&&=\frac{1}{2}~\frac{1}{\left[{\bf q}^2+(m+M({\bf q}))^2\right]^{1/2}}~\nonumber
\end{eqnarray}
we have
\begin{equation}
\label{39}
M({\bf p})=2G'N_c~\int \frac{d {\bf q}}{(2\pi)^3}~I({\bf p}-{\bf q})~
\frac{m+M({\bf q})}{\left[{\bf q}^2+(m+M({\bf q}))^2\right]^{1/2}}~.
\end{equation}
With the correlator corresponding to the NJL model we obtain the well
known gap equation
$$M=2G'N_c~\int^\Lambda \frac{d {\bf q}}{(2\pi)^3}
\frac{m+M}{\left[{\bf q}^2+(m+M)^2\right]^{1/2}}~.$$
For the Keldysh model we have
$$M({\bf p})=2G'N_c~\frac{m+M({\bf p})}{\left[{\bf p}^2+
(m+M({\bf p}))^2\right]^{1/2}}~,$$
and remember that $I({\bf p})=(2\pi)^3 \delta({\bf p})$. Then it is easy to
understand that the solution can be presented as a function of $p(M)$ for
convenient handling.

In the Hartree-Fock-Bogolyubov approach the following sum over the colour
matrices is used $\sum^{N_c^{2}-1}_{a=1}t^a_{ij}t^a_{jk}=
\frac{N_c^2-1}{2N_c}~\delta_{ik}$ and then we have for the quark mass
\begin{equation}
\label{40}
M_\theta({\bf p})=2G'~\frac{N_c^{2}-1}{N_c}\int \frac{d {\bf
q}}{(2\pi)^3}~I({\bf p}-{\bf q})~
\frac{|{\bf q}|}{|q_4|}~\sin\theta(q)~.
\end{equation}
Comparing this expression to Eq.(\ref{39}) it becomes clear that the four-
fermion interaction constant acquires the small correction $\sim 1/N_c$ which is
rooted in the mean field approximation while the higher order terms in $N_c$ are
held. The patent formula for the Keldysh model can be simply received in the
chiral limit. In the mean field approximation we have
$$M(p)=\left[(2G'N_c)^2-p^2\right]^{1/2}~$$
and in the Hartree-Fock-Bogolyubov we receive
$$M(p)=\left[\left(2G'~\frac{N_c^2-1}{N_c}\right)^2-p^2\right]^{1/2}~,$$
see also Eq.(\ref{34}). At $m\neq 0$ the momentum ~dependencies of masses are
quite different. For the Keldysh model in the mean field approximation at zero
momentum, for example, we have $M(0)=2G'N_c$ whereas in the Hartree-Fo approach
$M_\theta(0)=0$. At large momenta the mass in the mean field approximation behaves
as $|M(p)|\to 2G'N_c ~m/p$ and in the Hartree-Fock-Bogolyubov approximation it is
the following $|M(p)|\to \left(2G'~\frac{N_c^2-1}{N_c}\right)^2 m/p^2$,
see also Fig. \ref{f6}. However, generally, if one takes an orientation to the
analysis of integral characteristics $M(p)$ the results are not so different.
The similar relations could be obtained for the NJL model as well. Apparently it is
reasonable to notice here that our analysis of the Hamiltonian Eq.(\ref{9})
(Lagrangians Eqs.(\ref{36}), (\ref{377})) is also valid for the
Lorentz-invariant formulation when the $\tau$ ('time') integration is performed
for the quark fields as well (for the Lagrangians (\ref{36}) and (\ref{377})
the integration over 'time' is retained and the formfactor becomes a function of
four-vector $I(x-y)$).
\begin{equation}
\label{ad1}
{\cal L}=\bar q~(i \gamma_\mu \partial_\mu+im)~q-G\int d y~I(x-y)
~\langle \bar q~q'\rangle~\bar q'~q~.
\end{equation}
The selfconsistency condition Eq.(\ref{38})  acquires the covariant form. In
particular, in the Keldysh model in four-dimensional formulation when
$I(p)=(2\pi)^4~\delta(p)$ the mass gap
equation reads as $$M=4G N_c~\frac{m+M}{p^2+(m+M)^2}~.$$
It allows us to conclude that a quark never comes on the mass shell because
$$p^2+(m+M)^2=4 N_c~G~\frac{m+M}{M}>0~.$$
This feature has already been noticed in Ref. \cite{MZ2}. The similar behaviour
has been also observed in the analytic models of confinement \cite{ned}.
Meanwhile, an absence of bound states in the four-dimensional Keldysh model
(unlike the Keldysh model with three-dimensional formfactors)
is its shortage. There appears the additional integration over the fourth
component of auxiliary four-momentum $l$ in Eq. (\ref{26})
$$\int \frac{d {\bf q}}{(2\pi)^3} \to \int \frac{d l_4}{2\pi}~\int \frac{d{\bf
q}}{(2\pi)^3}~I(l_4)~\frac{1}{|p_4|+|q_4|-i~l_4}~,$$
where $I(l_4)$ is the respective part of the formfactor. In particular, for the
four-dimensional Keldysh model with $I(l_4)=2\pi~\delta(l_4)$ the mean energy
functional can be presented in the following way
\begin{widetext}
\begin{equation}
\label{314}
W(m)=\int \frac{d{\bf p}}{(2\pi)^3}~\left[|p_4|~\left(1-\cos\theta\right)-
G~\frac{p^2}{|p_4|^2}~\frac{1}{2|p_4|}\left(\sin\theta-\frac{m}{p}\cos\theta
\right)^2\right]~.
\end{equation}
\end{widetext}
The singularity revealed in three-dimensional Keldysh model manifests itself as
weaker (logarithmic only) one in the four-dimensional consideration.


\bibliography{apssamp}

\begin{thebibliography}{99}
\bibitem{ss}
T. Sch\"afer and E. V. Shuryak, Rev. of Mod. Phys. {\bf 70}, 323 (1998).
\bibitem{4}
M. K. Volkov and A. E. Radzhabov, Phys. Uspekhi. {\bf 176}, 569 (2006);\\
D. Ebert, H. Reinhardt, and M. K. Volkov, Prog. Part.
Nucl. Phys. {\bf 33}, 1 (1994).
\bibitem{NJL}
Y. Nambu and G. Jona-Lasinio, Phys. Rev. {\bf 122}, 345 (1961).
\bibitem{CDG}
C. G. Callan, R. Dashen, and D. J. Gross, Phys. Lett. {\bf 66}, 375 (1977);\\
C. G. Callan, R. Dashen, and D. J. Gross, Phys. Rev. {\bf D17}, 2717 (1978).
\bibitem{cooled}
A. Di Giacomo, E. Meggiolaro, and H. Panagopoulos,
Nucl. Phys. {\bf B483}, 271 (1997);\\
M. D\'Elia, A. Di Giacomo, and E. Meggiolaro,
Phys. Lett. {\bf B408}, 315 (1997);\\
G. Bali, N. Brambilla, and A. Vairo, Phys. Lett. {\bf B421}, 265 (1998);\\
M. D\'Elia, A. Di Giacomo, and E. Meggiolaro, Phys. Rev. {\bf D67}, 114504
(2003).
\bibitem{lattice}
A. E. Dorokhov, S. V. Esaibegyan, and S. V. Mikhailov, Phys. Rev. {\bf D56},
4062 (1997);\\
E.-M. Ilgenfritz, B. V. Martemyanov, S. V. Molodtsov, M. M\"uller-Preussker, and
Yu. A. Simonov, Phys. Rev. {\bf D58}, 114508 (1998);\\
E.-M. Ilgenfritz, B. V. Martemyanov, and M. M\"uller-Preussker,
Phys. Rev. {\bf D62},  096004 (2000).
\bibitem{DiakPetr}
D. I. Diakonov and V. Yu. Petrov, Nucl. Phys. {\bf B245}, 259 (1984).
\bibitem{Faddeev}
L. D. Faddeev, hep-th 0805.1624;\\
L. D. Faddeev and A. J. Niemi, Phys. Rev. Lett. {\bf 82}, 1624 (1999);\\
L. D. Faddeev and A. J. Niemi, Phys. Lett. {\bf B449}, 214 (1999);
{\bf B464}, 90 (1999).
\bibitem{vanBaal}
P. van Baal, Nucl. Phys. Proc. Suppl. {\bf B108}, 3 (2002); hep-th 01099148;\\
P. van Baal and A. Wipf, Phys. Lett. {\bf B515}, 181 (2001);\\
E. T. Tomboulis,  PoSLAT2007:336,2007, hep-lat 0712.2620;\\
K. R. Ito and E. Seiler, hep-lat 0711.4930.
\bibitem{Protog}
A. P. Protogenov, Phys. Uspekhi.  {\bf 176}, 689 (2006).
\bibitem{MolZin}
S. V. Molodtsov and G. M. Zinovjev, JHEP, 122, 112 (2008).
\bibitem{GtH}
G. 't Hooft, Phys. Rev. {\bf D14}, 3432 (1976); Phys. Rev. Lett. {\bf 37}, 8
(1976); Phys. Rep. {\bf 142}, 357 (1986)
\bibitem{DPSb}
D. I. Diakonov and V. Yu. Petrov, Nucl. Phys. {\bf B272}, 457 (1986);\\
D. I. Diakonov and V. Yu. Petrov, in {\it 'Hadronic Matter under Extreme
Conditions'}, ed. by V. Shelest and G. Zinovjev
(Naukova Dumka, Kiev, 1986), p. 192
\bibitem{Pobyl}
P. V. Pobylitsa, Phys. Lett. {\bf B226}, 387 (1989).
\bibitem{DS}
U. Marguard and H. G. Dosch, Phys. Rev. {\bf D35}, 2238 (1987);\\
H. G. Dosch, Phys. Lett. {\bf B190}, 177 (1987);\\
H. G. Dosch and Yu. A. Simonov, Phys. Lett. {\bf B205}, 339 (1988);
Z. Phys. {\bf C45}, 147 (1989).
\bibitem{2}
N. G. Van Kampen, Phys. Rep. {\bf 24}, 171 (1976);
Physica {\bf 74}, 215, 239 (1974).
\bibitem{Ker}
B. O. Kerbikov, D. S. Kuzmenko, and Yu. A. Simonov,
JETP Lett. {\bf 65}, 137 (1997);\\
A.G. Zubkov, O.V. Dubasov, and B.O. Kerbikov,
Int. J. Mod. Phys. {\bf A 14} (1999) 241.
\bibitem{Simon}
Yu. A. Simonov, Phys. Lett. {\bf B412}, 371 (1997).
\bibitem{1}
R. P. Feynman and F. L. Vernon, Jr., Ann. Phys. {\bf 24}, 118 (1963);\\
Feynman, Richard P. and Hibbs, A. R., 'Quantum Mechanics and Path Integrals', \\
McGraw-Hill, New York, 1965.
\bibitem{7}
A. Di Giacomo, H. D. Dosch, V. I. Shevchenko, and Yu. A. Simonov,
Phys. Rep. {\bf 372}, 319 (2002).
\bibitem{Bornyak}
E.-M. Ilgenfritz, M. M\"uller-Preussker, A. Sternbeck, A. Schiller,
and I. L. Bogolubsky, Braz.J.Phys. {\bf 37} (2007) 193;\\
I. L. Bogolubsky, V. G. Bornyakov, G. Burgio, E.-M. Ilgenfritz, M. M\"uller-
Preussker, and V. K. Mitrjushkin,
13th Lomonosov Conference on Elementary Particle Physics,
Moscow, August 2007, hep-lat/0804.1250.
\bibitem{shafer}
T. Sch\"afer, hep-lat/0411010.
\bibitem{Zah}
V. I. Zakharov, Phys. Uspekhi, {\bf 174}, 39 (2004).
\bibitem{shrink}
MILC Collaboration, C. Aubin et al., arXiv:0410024;\\
V. I. Zakharov, arXiv:061234.
\bibitem{mm}
M. V. Martemyanov, S. V. Molodtsov,
Pisma JETF, {\bf 65}, 133 (1997).
\bibitem{3}
N. N. Bogolyubov, Journal of Phys, {\bf 9}, 23 (1947).
\bibitem{Y}
A. Le Yaouanc, L. Oliver, O. P\'ene, and J.-C. Raynal,
Phys. Rev. {\bf D29}, 1233 (1984);\\
A. Le Yaouanc, L. Oliver, S. Ono, O. P\'ene, and J.-C. Raynal,
Phys. Rev. {\bf D31}, 137 (1985).
\bibitem{kriv}
M. I. Krivoruchenko, Phys. Uspekhi, {\bf 164}, 643 (1994),
Yad. Fiz. {\bf 47}, 1823 (1987).
\bibitem{5}
T. Hatsuda and T. Kunihiro, Phys. Rep. {\bf 247}, 221 (1994).
\bibitem{6}
L. V. Keldysh, Doctor thesis (FIAN, 1965) (unpublished);\\
E. V. Kane, Phys. Rev. {\bf 131}, 79 (1963);\\
V. L. Bonch-Bruevich, in 'Physics of solid states', M., VINITI, 1965.
\bibitem{9}
B. A. Arbuzov, M. K. Volkov, and I. V. Zaitsev, Int. J. of Mod. Phys. {\bf A21},
5721
(2006);\\
B. A. Arbuzov, Phys. Lett. {\bf B656}, 67 (2007).
\bibitem{weAr}
S. V. Molodtsov and G. M. Zinovjev, arXiv:0811.4405.
\bibitem{MVZ}
S. V. Molodtsov, M. K. Volkov, and G. M. Zinovjev, Teor. Mat. Fiz. (in print),
hep-ph 0812.2666.
\bibitem{stoch}
B. V. Chirikov, Phys. Rep. {\bf 52}, 263 (1979);\\
G. M. Zaslavsky, Phys. Rep. {\bf 80}, 157 (1981); \\
UJFG, NATO ASI, Les Houches, session LII, 1-31 Aout, 1989,
Q172.5.C45C429 {\it 'Chaos and Quantum Physics'},
ed. by M.-J. Jiannony, A. Voros, et J. Zinn-Zustin,  1991;\\
W. H. Zurek, Rev. Mod. Phys. {\bf 75}, 715 (2003).
\bibitem{MZ2}
A. E. Dorokhov, G. M. Zinovjev, and S. V. Molodtsov,
Yad. Fiz. {\bf 71}, 785 (2008).
\bibitem{ned}
G. V. Efimov and S. N. Nedelko, Eur. Phys. J. {\bf C1}, 343 (1998);\\
A. C. Kalloniatis and S. N. Nedelko, Phys. Rev. {\bf D64}, 114025 (2001).
\end{thebibliography}

\end{document}